\documentclass[pre,twocolumn,amsmath,amssymb,nofootinbib,floatfix,superscriptaddress]{revtex4}

\usepackage{graphicx,bm}
\usepackage[normalem]{ulem} 

\makeatletter
\def\graphicscale{\twocolumn@sw{0.3}{0.4}}
\def\graphicthreescale{\twocolumn@sw{0.3}{0.4}}

\begin{document}

\title{Lattice Abelian-Higgs model with noncompact
gauge fields}

\author{Claudio Bonati} 
\affiliation{Dipartimento di Fisica dell'Universit\`a di Pisa
        and INFN Largo Pontecorvo 3, I-56127 Pisa, Italy}

\author{Andrea Pelissetto}
\affiliation{Dipartimento di Fisica dell'Universit\`a di Roma Sapienza
        and INFN Sezione di Roma I, I-00185 Roma, Italy}

\author{Ettore Vicari} 
\affiliation{Dipartimento di Fisica dell'Universit\`a di Pisa
        and INFN Largo Pontecorvo 3, I-56127 Pisa, Italy}

\date{\today}

\begin{abstract}

We consider a noncompact lattice formulation of the three-dimensional
electrodynamics with $N$-component complex scalar fields, i.e., the
lattice Abelian-Higgs model with noncompact gauge fields. For any
$N\ge 2$, the phase diagram shows three phases differing for the
behavior of the scalar-field and gauge-field correlations: the Coulomb
phase (short-ranged scalar and long-ranged gauge correlations), the
Higgs phase (condensed scalar-field and gapped gauge correlations),
and the molecular phase (condensed scalar-field and long-ranged gauge
correlations).  They are separated by three transition lines meeting
at a multicritical point. Their nature depends on the coexisting
phases and on the number $N$ of components of the scalar field. In
particular, the Coulomb-to-molecular transition line (where gauge
correlations are irrelevant) is associated with the
Landau-Ginzburg-Wilson $\Phi^4$ theory sharing the same SU($N$) global
symmetry but without explicit gauge fields. On the other hand, the
Coulomb-to-Higgs transition line (where gauge correlations are
relevant) turns out to be described by the continuum Abelian-Higgs
field theory with explicit gauge fields.  Our numerical study is based
on finite-size scaling analyses of Monte Carlo simulations with $C^*$
boundary conditions (appropriate for lattice systems with noncompact
gauge variables, unlike periodic boundary conditions), for several
values of $N$, i.e., $N=2, 4, 10, 15$, and $25$.  The numerical
results agree with the renormalization-group predictions of the
continuum field theories.  In particular, the Coulomb-to-Higgs
transitions are continuous for $N\gtrsim 10$, in agreement with the
predictions of the Abelian-Higgs field theory.
 
\end{abstract}

\maketitle


\section{Introduction}
\label{intro}

Models of scalar fields with U(1) gauge symmetry and SU($N$) global
symmetry emerge as effective theories of superconductors, superfluids,
and of quantum SU($N$) antiferromagnets ~\cite{RS-90, TIM-05, TIM-06,
  Kaul-12, KS-12, BMK-13, NCSOS-15, WNMXS-17}.  In particular,
three-dimensional (3D) classical U(1) gauge models with $N=2$
supposedly describe the transition between the N\'eel and the
valence-bond-solid state in two-dimensional antiferromagnetic SU(2)
quantum systems~\cite{Sandvik-07, MK-08, JNCW-08, Sandvik-10,
  HSOMLWTK-13, CHDKPS-13, PDA-13, SGS-16}, that represent the
paradigmatic models for the so-called deconfined quantum criticality
\cite{SBSVF-04}.

This class of quantum models and their classical counterparts have
been extensively studied with the purpose of identifying the nature of
their different phases and transitions.  It has been realized that
topological aspects, like the Berry phase or the compact/noncompact
nature of the gauge fields, play a crucial role in determining the
nature of the transition.  For example, the critical behavior of the
simplest classical model with U(1) gauge symmetry, the lattice
CP$^{N-1}$ model, drastically depends on the presence/absence of
topological defects~\cite{PV-20-mfcp,MV-04,MS-90}, such as monopoles,
both for large and small values of $N$, in particular for $N=2$.
Analogous differences emerge in the behavior of compact and noncompact
lattice formulations of scalar electrodynamics, i.e., of the
multicomponent Abelian-Higgs (AH) model. In particular, for $N=2$,
theoretical and numerical investigations of classical and quantum
transitions that are expected to be in the same universality class as
those occurring in noncompact scalar electrodynamics have provided
evidence of weakly first-order or continuous transitions belonging to
a new universality class, see, e.g., Refs.~\cite{SBSVF-04, KPST-06,
  Sandvik-07, MK-08, JNCW-08, MV-08, KMPST-08-a, KMPST-08, CAP-08,
  LSK-09, CGTAB-09, CA-10, BDA-10, Sandvik-10, Bartosch-13,
  HSOMLWTK-13, CHDKPS-13, PDA-13, BS-13, NCSOS-15, NSCOS-15, SP-15,
  SGS-16, PV-20-mfcp, SN-19, SZ-20}.

\begin{figure}[tbp]
\includegraphics*[width=0.8\columnwidth]{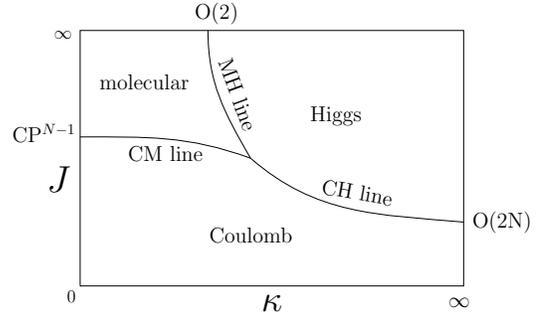}
  \caption{Sketch of the phase diagram of the lattice AH model with
    noncompact gauge fields and unit-length $N$-component complex
    scalar fields, for generic $N\ge 2$.  Three transition lines can
    be identified: the Coulomb-to-Higgs (CH) line between the Coulomb
    and Higgs phases, the Coulomb-to-molecular (CM) line, and the
    molecular-to-Higgs (MH) line.  They are continuous or of first
    order depending on the value of $N$, see Sec.~\ref{phadia} for
    details.  We also report the models emerging in some limiting
    cases: the CP$^{N-1}$ model for $\kappa=0$, the O($2N$) vector
    model for $\kappa\to\infty$, and the inverted XY or O(2) model for
    $J\to\infty$.  }
\label{phdiasketch}
\end{figure}

Here we present a numerical study of the phase diagram of the AH model
with noncompact gauge fields, for several values of $N$, the number of
components of the complex scalar field.  Our study confirms the
existence of important differences with the compact AH
model~\cite{PV-19-AH3d}, for both small and large values of $N$.

In Fig.~\ref{phdiasketch} we sketch the phase diagram of the
noncompact lattice AH model with unit-length $N$-component scalar
fields. For any $N\ge 2$ the phase diagram is characterized by three
phases. There is a Coulomb phase, in which the global SU($N$) symmetry
is unbroken and the electromagnetic correlations are long-ranged. The
other two phases are characterized by the breaking of the SU($N$)
symmetry. They are distinguished by the behavior of the gauge
modes. In the Higgs phase, electromagnetic correlations are gapped,
while in the molecular phase the electromagnetic field is ungapped.
The Coulomb, molecular, and Higgs  phases are separated by three different transition lines meeting
at one point of the phase diagram.  The nature of the transition lines
is different, due to the fact that they separate different phases.
Moreover, their nature crucially depends on the number $N$ of
components.

Our numerical study allows us to characterize the nature of the
different transition lines.  For large $N$, the critical behavior
along the Coulomb-to-Higgs (CH) transition line belongs to the
universality class associated with the stable fixed point of the
multicomponent AH field theory, which predicts a continuous transition
only for a large number of components (we present numerical evidence
of continuous transitions for $N\ge 10$), and in particular in the
large-$N$ limit. For small $N$, instead, the CH line is characterized
by weak first-order transitions (this is the case for $N=2, 4$).
Along the Coulomb-to-molecular (CM) transition line, gauge
correlations do not play any particular role. Numerical results are
consistent with the predictions of the Landau-Ginzburg-Wilson (LGW)
$\Phi^4$ field theory with SU($N$) global symmetry and without
explicit gauge fields.  Therefore, for $N=2$ CM transitions are
continuous and belong to the O(3) vector universality class, while
they are expected to be of first order for any larger value of $N$.
Finally, the molecular-to-Higgs (MH) transitions are essentially
related to the behavior of the gauge correlations. They are expected
to belong to the universality class of the inverted XY model, for any
$N$.

The paper is organized as follows.  In Sec.~\ref{models} we introduce
the lattice AH model. In Sec.~\ref{FTanalysis} we discuss the
field-theoretical models that may be relevant for the phase
transitions of the model.  In Sec.~\ref{phadia} we present the
possible scenarios for the phase diagram and for the nature of the
transition lines, focussing on some limits of the Hamiltonian
parameters. Sec.~\ref{numres} presents our numerical results, for
$N=2, 4, 10, 15$ and $N=25$.  Finally, we summarize and draw our
conclusions in Sec.~\ref{conclu}.  In the appendix we discuss the
pathologies of periodic boundary conditions in systems with noncompact
gauge variables (averages of gauge-invariant noncompact Polyakov lines
are not defined); to overcome this problem, we introduce $C^*$
boundary conditions \cite{KW-91,LPRT-16}, which allow a rigorous
definition of the model in a finite volume.

\section{Three-dimensional lattice Abelian-Higgs models}
\label{models}

We consider $d$-dimensional systems characterized by a global SU($N$)
symmetry and a local U(1) gauge symmetry. A paradigmatic quantum field
theory with these symmetries is the multicomponent scalar
electrodynamics, or AH field theory, in which an $N$-component complex
scalar field ${\bm \Phi}$ is minimally coupled to the electromagnetic
field $A_\mu$. The corresponding continuum Lagrangian reads
\begin{equation}
{\cal L} = 
|D_\mu{\bm\Phi}|^2
+ r\, {\bm \Phi}^*{\bm \Phi} + 
\frac{1}{6} u \,({\bm \Phi}^*{\bm \Phi})^2 + 
\frac{1}{4 g^2} \,F_{\mu\nu}^2 
\,,
\label{abhim}
\end{equation}
where $F_{\mu\nu}\equiv \partial_\mu A_\nu - \partial_\nu A_\mu$, and
$D_\mu \equiv \partial_\mu + i A_\mu$.

In the following we consider lattice models that are formal
discretizations of the continuum AH model. In particular, one may
consider lattice models that differ on the topological---compact or
noncompact---nature of the gauge fields.  We consider unit-length
$N$-component complex variables ${\bm z}_{\bm x}$ associated with each
site of a cubic lattice and gauge fields associated with the lattice
links.  The lattice Hamiltonian reads
\begin{eqnarray}
H &=& H_z + H_g \,,\label{ahham}
\end{eqnarray}
with
\begin{equation}
H_z = - J N \sum_{{\bm x}, \mu} 2\, {\rm Re}\,(\bar{\bm{z}}_{\bm x} \cdot
\lambda_{{\bm x},\mu}\, {\bm z}_{{\bm x}+\hat\mu}) \,,
\label{hz}
\end{equation}
where the sum runs over all links of the cubic lattice, and
$\lambda_{{\bm x},\mu}$ is a complex gauge field with $|\lambda_{{\bm
    x},\mu}| = 1$. In compact formulations the link phase
$\lambda_{{\bm x},\mu}$ is the fundamental gauge variable. The
corresponding simplest gauge Hamiltonian reads
\begin{eqnarray}
&&H_g = -  \kappa \sum_{{\bm x},\mu>\nu} {\rm Re}\,
(\lambda_{{\bm x},{\mu}} \,\lambda_{{\bm x}+\hat{\mu},{\nu}} 
\,\bar{\lambda}_{{\bm x}+\hat{\nu},{\mu}}  
  \,\bar{\lambda}_{{\bm x},{\nu}}) 
\,,\qquad \label{chg}
\end{eqnarray}
where the sum is over the lattice plaquettes and $\kappa$ plays the
role of inverse gauge coupling.  The partition function is $Z =
\sum_{\{{\bm z},\lambda\}} e^{-\beta H}$.

In noncompact formulations the fundamental gauge variable is the real
vector field $A_{{\bm x},\mu}$ and
\begin{equation}
\lambda_{{\bm x},\mu} = e^{iA_{{\bm x},\mu}}\,.
\label{nclink}
\end{equation}
In this case, the gauge Hamiltonian $H_g$ can be straightforwardly
derived from the continuum theory (\ref{abhim}), by replacing the
tensor field $F_{\mu\nu}({\bm x})$ with its discretized lattice
counterpart, i.e.
\begin{equation}
H_g = {\kappa\over 2} \sum_{{\bm x},\mu>\nu} 
(\Delta_{\hat\mu} A_{{\bm x},\nu} - 
\Delta_{\hat\nu} A_{{\bm x},\mu})^2\,.
\label{nchg}
\end{equation}
Here the sum runs over all plaquettes, $\Delta_{\hat\mu}$ denotes the
discretized derivative along $\hat\mu$ (i.e. $\Delta_{\hat\mu} A_{\bm
  x} \equiv A_{{\bm x}+\hat\mu} - A_{\bm x}$), and $\kappa\ge 0$
corresponds to the inverse gauge coupling~$1/g^2$ of the continuum
theory (\ref{abhim}).  The partition function reads
\begin{equation}
Z = \sum_{\{{\bm z},A\}} e^{-\beta H}\,.
\label{ncz}
\end{equation}
In the following we rescale $J$ and $\kappa$ by $\beta$, thus formally
setting $\beta=1$.

It is important to note that, at variance with the compact case, the
partition function $(\ref{ncz})$ is only formally defined.  Because of
gauge invariance, there is an infinite number of zero modes, therefore $Z =
\infty$.  As discussed in detail in App.~\ref{appcstar}, by an
appropriate choice of boundary conditions and by restricting our
attention to gauge-invariant observables, we can make $Z$, as well as
any gauge-invariant average, well-defined. This is, of course, of
crucial importance for the numerical computation. Note that periodic
boundary conditions cannot be used for the noncompact model. Indeed,
in this case the Polyakov loops in terms of the noncompact variables
are not bounded and never thermalize; thus, even gauge-invariant
observables are ill-defined.

In the following we study the phase diagram and the transition lines
of the noncompact model (\ref{nchg}). Appropriate order parameters can
be defined in terms of the ${\bm z}_{\bm x}$ and $A_{{\bm x}, \mu}$
fields. In our study we focus on the correlations of the
gauge-invariant bilinear operator
\begin{equation}
Q_{{\bm x}}^{ab} = \bar{z}_{\bm x}^a z_{\bm x}^b - {1\over N}
\delta^{ab}\,,
\label{qdef}
\end{equation}
which is a hermitian and traceless $N\times N$ matrix that transforms
as $Q_{{\bm x}} \to {U}^\dagger Q_{{\bm x}} \,{U}$ under the global
SU($N$) transformations.

\section{Field-theoretical approaches}
\label{FTanalysis}

One of the motivations of this work is that of understanding the
relation between the phase diagram of 3D lattice Abelian gauge models
and the renormalization-group (RG) flow of the continuum AH model
(\ref{abhim}), which has been studied within the $\varepsilon\equiv
4-d$ expansion framework~\cite{HLM-74,FH-96,IZMHS-19}, using the
functional RG approach ~\cite{FH-17}, and in the large-$N$
limit~\cite{HLM-74,DHMNP-81,YKK-96,MZ-03,KS-08}.  One expects that the
3D RG flow of the continuum AH model describes some critical
transitions occurring in 3D statistical systems characterized by an
Abelian gauge symmetry and a global SU($N$) symmetry. However, as far
as we know, the correspondence between the transition lines observed
in lattice systems and the fixed points of the continuum AH model has
not been fully clarified yet.

\subsection{RG flow of the AH field theory}
\label{epsexp}

In the $\varepsilon$-expansion framework, the RG flow is determined by
the $\beta$ functions associated with the renormalized couplings $u$
and $f\equiv g^2$. One-loop computations give~\cite{HLM-74}
\begin{eqnarray}
 &&\beta_u \equiv \mu {\partial u\over \partial \mu}
 = - \varepsilon u + 
 (N+4) u^2 - 18 u f + 54 f^2\,,
 \nonumber \\
 &&\beta_f \equiv \mu {\partial f \over \partial \mu}
 = - \varepsilon f + N f^2\,
 \label{betafunc}
 \end{eqnarray} 
[we used rescaled couplings $u\to u/(24\pi^2)$ and $f\to f/(24\pi^2)$
  to simplify the equations].  A stable fixed point is present only
for
\begin{equation}
N \geq N_{4} = 90 + 24\sqrt{15} \approx 183\, .
\label{N4}
\end{equation}
It is located at
\begin{eqnarray}
u^*  =  {N+18 + \sqrt{N^2-180N-540}\over 2 N(N+4)}\,\varepsilon\,,
\;\; f^* = {\varepsilon \over N}\,.\quad
\label{fipo}
\end{eqnarray}
More generally, in generic dimensions $d = 4 -\varepsilon$, a stable
fixed point exists only for $N > N_c(\varepsilon)$.  This implies that
3D lattice AH models may undergo a continuous transition associated
with the AH stable fixed point only if
\begin{equation}
N>N_c\equiv N_c(1)\,.  
\label{ncdef}
\end{equation}
The critical number of components $N_c(\varepsilon)$ has been
determined to four loops \cite{IZMHS-19}:
\begin{equation}
N_c(\varepsilon) = N_{4}\left[1 - 1.752 \,\varepsilon + 0.789\,
\varepsilon^2 + 0.362 \,\varepsilon^3+O(\varepsilon^4)\right]  .
\label{nceps}
\end{equation}
The large coefficients of the expansion (\ref{nceps}) make a reliable
3D (i.e., for $\varepsilon=1$) estimate quite problematic.
Nevertheless, by means of a resummation of the expansion that takes
somehow into account two-dimensional results, Ref.~\cite{IZMHS-19}
obtained $N_c = 12.2(3.9)$ in three dimensions, which confirms the
absence of a stable fixed point for small values of $N$.

In the limit $\kappa \to\infty$, the gauge fields order so that
$\lambda_{{\bm x},\mu} = 1$. The lattice AH model becomes equivalent
to the symmetric O$(2N)$ vector theory. Therefore, for large $\kappa$,
one expects significant crossover effects, which increase as $\kappa$
increases, due to the nearby O$(2N)$ critical transition.  In the
continuum AH model, the crossover is controlled by the RG flow in the
vicinity of the O$(2N)$ fixed point
\begin{equation}
u^*_{{\rm O}(2N)}= {1\over N+4}  \varepsilon\,,\qquad
f=0\,.
\label{O2nfp}
\end{equation}
This fixed point exists for any $N$ and is always unstable.  The
analysis of the stability matrix $\Omega_{ij} = \partial
\beta_i/\partial g_j$ shows that it has a positive eigenvalue
$\lambda_u=\omega$, where $\omega>0$ is the exponent controlling the
leading scaling corrections in O$(2N)$ vector models~\cite{PV-02}, and
a negative eigenvalue $\lambda_f$, which makes the fixed point
unstable.  Since~\cite{PV-19-AH3d} $\lambda_f=-\varepsilon$ to all
orders in perturbation theory, the RG dimension $y_f=-\lambda_f =
\varepsilon$ of the operator that controls the crossover behavior is
one in three dimensions.

\subsection{The AH field theory  in the large-$N$ limit}
\label{largen}

The existence of a stable fixed point for sufficiently large values of
$N$ and, therefore, of a universality class described by the AH field
theory, is confirmed by $1/N$
calculations~\cite{HLM-74,YKK-96,MZ-03,KS-08}.  Critical exponents
have also been computed \cite{HLM-74,YKK-96} to order $1/N$.  For the
critical exponent $\nu$ associated with the correlation length, one
finds~\cite{HLM-74}
\begin{eqnarray}
\nu = 1 - \frac{48}{\pi^2 N} + O(N^{-2})\,,
\label{nulargen}
\end{eqnarray} 
for the three-dimensional model.
Also the critical behavior of the two-point function 
\begin{equation}
G({\bm
x},{\bm y}) = \langle {\rm Tr}\, B({\bm x}) B({\bm y}) \rangle
\label{gft}
\end{equation}
of the gauge-invariant bilinear composite operator
\begin{equation}
B_{ab}({\bm x}) = \Phi_{a}({\bm x})^\dagger 
\Phi_b({\bm x})-{1\over N}\delta_{ab}
|\Phi|^2\,
\label{Bdef}
\end{equation}
has been considered.  Here, $ B_{ab}({\bm x})$ is the coarse-grained
continuum counterpart of the lattice operator $Q^{ab}_{{\bm x}}$
defined in Eq.~(\ref{qdef}).  At the critical point, $G({\bm x},{\bm
  y})$ has the power-law behavior
\begin{equation}
G({\bm x},{\bm y})|_{J=J_c}\sim \frac{1}{|{\bm x}-{\bm  y}|^{d-2+\eta_q}}
\label{etaqdef}
\end{equation}  
characterized by the critical exponent $\eta_q$.  At order $1/N$ one
finds~\cite{YKK-96}
\begin{equation}
\eta_q = 1 - \frac{32}{\pi^2 N}  + O(N^{-2})\,,
\label{etalargen}
\end{equation}
in three dimensions.

\subsection{The gauge-invariant LGW  framework}
\label{GLW}

A second approach that has been used to predict the critical behavior
of lattice AH models \cite{PV-19-AH3d,PV-19-CP} is the LGW
framework~\cite{WK-74,Fisher-75,PV-02}.  It assumes that the relevant
critical modes are associated with the gauge-invariant local site
variable (\ref{qdef}).  As discussed in
Refs.~\cite{PTV-17,PV-19-AH3d,PV-19-CP,BPV-19-sqcd,BPV-19-son}, this
is a highly nontrivial assumption, as it postulates that gauge fields
do not play a relevant role in the effective theory. In this approach,
the order-parameter field is a traceless hermitian matrix field
$\Psi^{ab}({\bm x})$, which can be formally defined as the average of
$Q_{\bm x}^{ab}$ over a large but finite lattice domain.  The LGW
field theory is obtained by considering the most general fourth-order
polynomial in $\Psi$ consistent with the SU($N$) global symmetry:
\begin{eqnarray}
{\cal H}_{\rm LGW} &=& {\rm Tr} (\partial_\mu \Psi)^2 
+ r \,{\rm Tr} \,\Psi^2 \label{hlg}\\
&+&   w \,{\rm tr} \,\Psi^3 
+  \,u\, ({\rm Tr} \,\Psi^2)^2  + v\, {\rm Tr}\, \Psi^4 .
\nonumber
\end{eqnarray}
Also in this framework continuous transitions may only arise if the RG
flow in the LGW theory has a stable fixed point.

For $N=2$, the cubic term in Eq.~(\ref{hlg}) vanishes and the two
quartic terms are equivalent.  Therefore, one recovers the
O(3)-symmetric vector LGW theory.  Thus, for $N=2$ continuous
transitions, belonging to the Heisenberg universality class, are
possible.  For $N\ge 3$, the cubic term is generically present.  Its
presence is usually taken as an indication that phase transitions
occurring in this class of systems are generally of first order.
Indeed, a straightforward mean-field analysis shows that the
transition is of first order in four dimensions, where the mean field
approximation is exact.  If statistical fluctuations are small---this
is the basic assumption---the transition is of first order also in
three dimensions.  In this approach, continuous transitions may still
occur, but they require a fine tuning of the microscopic parameters,
leading to the effective cancellation of the cubic term~\cite{DPV-15}.

It is important to note that the field-theoretical approaches based on
the continuum AH field theory (\ref{abhim}) and the effective LGW
field theory (\ref{hlg}) are not equivalent, as they make different
assumptions on the role of the gauge correlations. They give different
predictions, both for small and large values of $N$.  For $N=2$ the
continuum AH model predicts the absence of continuous transitions, due
to the absence of a stable fixed point. On the other hand, a stable
O(3) vector fixed point exists in the effective LGW theory, leaving
open the possibility of observing continuous transitions.

For large values of $N$ (more precisely, for $N > N_c$, see
Sec.~\ref{epsexp}), the continuum AH theory and the effective LGW
approach give again contradictory results. If one trusts the argument
based on the relevance of the cubic term, the LGW approach predicts a
first-order transition unless a fine tuning of the microscopic
parameters is performed. Instead, continuous transitions are possible
without any fine tuning according to the continuum AH field theory.
For intermediate values of $N$, that is for $3 \le N < N_c$, both
approaches predict lattice models to undergo first-order transitions.

The field-theoretical predictions have been compared with numerical
results for the lattice CP$^{N-1}$ and AH models, loop models, and 2D
quantum systems
\cite{NCSOS-11,NCSOS-13,PV-20-largeN,NCSOS-15,KS-12,BMK-13}.
Simulations of the CP$^1$ model~\cite{PV-19-CP,PV-19-AH3d} confirm the
existence of continuous transitions, belonging to the O(3) vector
universality class, for $N=2$.  Therefore, the LGW theory provides the
correct description of the large-scale behavior of these
systems. Evidently, gauge correlations are not relevant in the CP$^1$
model, and the continuum AH model does not predict the correct
behavior.

For $N=3$ numerical results are not yet conclusive.  Indeed numerical
simulations for the lattice CP$^2$ model favor a first order
transition~\cite{PV-19-CP}, while the results for the loop model of
Refs.~\cite{NCSOS-11,NCSOS-13} apparently favor a continuous
transition.  The available numerical results for lattice CP$^{N-1}$
models for $N\ge 4$ are consistent with first-order
transitions~\cite{NCSOS-13,PV-19-CP,PV-20-largeN,KHI-11}, again
confirming the LGW predictions.  We note that a continuous transition
has been observed in a monopole-free version of the CP$^{N-1}$ model
\cite{PV-20-mfcp} for large values of $N$, demonstrating the relevance
of these topological defects.

\section{The phase diagram of the noncompact lattice AH model}
\label{phadia}

\subsection{Transition lines and limiting cases}
\label{limcases}

To sketch the phase diagram of the noncompact lattice AH model with
$N$-component scalar fields, see Fig.~\ref{phdiasketch}, it is useful
to consider some particular cases, in which the thermodynamic behavior
is already known. No transitions are expected along the $J=0$ line,
while transitions occur along the $\kappa=0$, the $J=\infty$, and the
$\kappa=\infty$ lines.

{\em Phase diagram along the $\kappa=0$ line.}  For $\kappa=0$ the
lattice AH model is equivalent to a lattice formulation of the
CP$^{N-1}$ models with explicit lattice gauge
variables~\cite{PV-19-CP}.  Its phase diagram has been recently
discussed in Refs.~\cite{PV-19-CP,PV-20-largeN}.  There are two phases
separated by a finite-temperature transition, where the order
parameter is the gauge-invariant matrix defined in Eq.~(\ref{qdef}).
The available estimates of the transition point $J_c$ are reported in
Table~\ref{tab:Jc}. The phase transition is continuous for $N=2$,
belonging to the O(3) vector universality class~\cite{PV-02} (accurate
estimates of the O(3) critical exponents can be found in
Refs.~\cite{Hasenbusch-20,KP-17,HV-11,CHPRV-02,GZ-98}), and of first
order for $N\ge 3$. It is very weak for $N=3$~\cite{PV-19-CP}, and it
becomes stronger and stronger with increasing $N$~\cite{PV-20-largeN}.
It is natural to conjecture that analogous transitions occur along the
CM line for small values of $\kappa$.  Thus, we expect a line of
continuous transitions belonging to the O(3) universality class for
$N=2$, and a line of first-order transitions for $N\ge 3$.  Note that,
for $N=2$, this scenario implies the stability of the O(3) critical
behavior against perturbations due to the noncompact gauge field.  We
shall report numerical evidence that confirms it.

{\em Phase diagram along the $J=\infty$ line.}  For $J\to\infty$ the
relevant configurations are those that minimize $H_z$,
cf. Eq.~(\ref{hz}).  There is no frustration, so that we obtain
\begin{equation} 
\bar{\bm{z}}_{\bm x} \cdot \lambda_{{\bm x},\mu}\, {\bm z}_{{\bm
    x}+\hat\mu} = 1\,,
\label{jinflim}
\end{equation}
which implies  ${\bm{z}}_{\bm x} = \lambda_{{\bm x},\mu}\, {\bm z}_{{\bm
x}+\hat\mu}$. A repeated use of this relation along a plaquette implies
\begin{equation}
\lambda_{{\bm x},{\mu}} \,\lambda_{{\bm x}+\hat{\mu},{\nu}}
\,\bar{\lambda}_{{\bm x}+\hat{\nu},{\mu}}
  \,\bar{\lambda}_{{\bm x},{\nu}} = 1
\end{equation}
on each plaquette. Therefore, by an appropriate gauge transformation
we obtain $A_{{\bm x}, \mu} = 2 \pi n_{{\bm x}, \mu}$, where $n_{{\bm
    x}, \mu} \in {\mathbb Z}$. We thus obtain a dual loop
representation of the 3D XY model, which is expected to undergo an
``inverted'' XY transition, i.e., a transition belonging to the XY
universality class but with inverted high and low temperature
phases~\cite{DH-81}.  Such a transition occurs at~\cite{NRR-03}
$\kappa_c(J=\infty) = 0.076051(2)$ [we obtained it by using the
  estimate $\beta_c=3.00239(6)$ reported in Ref.~\cite{NRR-03} and
  identifying $\kappa_c = \beta_c/(4\pi^2)$]. Note that $N$ does not
play any role here, thus the critical behavior does not depend on $N$.
It is natural to conjecture that the same behavior holds for finite
but large $J$. Thus, we expect a line of continuous transitions of the
inverted XY type separating the molecular and Higgs phases.

\begin{table}
\begin{tabular}{rclcclc}
\hline\hline 
$N$ &  $\qquad$ &
$J_c\, (\kappa=0)$  &   & $\quad$ &  $J_c \,(\kappa=\infty$) &  \\  
\hline
2  & &  0.7102(1)~\cite{PV-19-CP}  & O(3) && 
0.233965(2)~\cite{BFMM-96} &  O(4) \\
3   &&  0.6196(2)~\cite{PV-19-CP}  & FO  && 
0.23813(3)~\cite{BC-97} & O(6)  \\
4   &&  0.5636(1)~\cite{PV-19-CP}  & FO  && 
0.24084(1)~\cite{DPV-15}  & O(8) \\
7   && 0.4714(5)~\cite{PV-20-largeN}  & FO &&    
0.244  & O(14) \\
10   && 0.4253(5)~\cite{PV-20-largeN}  & FO &&
0.247    & O(20) \\
15   && 0.381(1)~\cite{PV-20-largeN}  & FO &&
0.249  & O(30) \\
20   && 0.353~\cite{PV-20-largeN}  & FO && 
0.250  & O(40) \\
$\infty$ && 0.252731... & FO && 0.252731...~\cite{CPRV-96}& O($\infty$)\\
\hline\hline
\end{tabular}
\caption{Estimates of the critical values $J_c$ for $\kappa=0$ and
  $\kappa\to\infty$. We also report the nature of the transition: FO
  and O($n$) indicate a first-order transition and a continuous
  transition in the O($n$) vector universality class, respectively.
  The estimates of $J_c$ for $\kappa\to\infty$ for $7\le N \le 20$ are
  obtained by interpolating the results of Ref.~\cite{CPRV-96} for the
  lattice O($n$) vector model (the uncertainty on these interpolations
  is safely below 1 on the last reported digit).  For $N\to \infty$,
  the results of Ref.~\cite{CPRV-96} allow us to obtain
  $J_c(\kappa\to\infty) = b_\infty + b_{\infty,1} N^{-1} + O(N^{-2})$,
  with $b_\infty=0.252731...$ and $b_{\infty,1}\approx -0.0585$. }
  \label{tab:Jc}
\end{table}

{\em Phase diagram along the $\kappa=\infty$ line.}  In the
$\kappa\to\infty$ limit, we have $A_{{\bm x},\mu}=0$ apart from gauge
transformations.  Therefore, in this limit the $N$-component AH model
can be exactly mapped onto the standard real $2N$-component vector
model, which undergoes a continuous transitions for any $N$, see
Table~\ref{tab:Jc}. At finite $\kappa$, the RG flow of the continuum
AH model predicts that gauge modes are a relevant perturbation of the
O($2N$) fixed point, see Eq.~(\ref{O2nfp}) and the subsequent
discussion in Sec.~\ref{epsexp}.  Therefore, if the CH transitions are
continuous, they do not belong to the O($2N$) vector universality
class.  However, the O($2N$) continuous transition for $\kappa=\infty$
may give rise to crossover phenomena for large values of $\kappa$.

On the basis of the above considerations, the most natural hypothesis
of phase diagram is the one reported in Fig.~\ref{phdiasketch}, with
three different phases.  For small $J$ and any $\kappa$ there is a
phase (it will be named Coulomb phase) in which the $z$ fields are
disordered and the gauge modes are in the inverted XY low-temperature
phase; for large $J$ and large $\kappa$ (Higgs superconducting phase)
there is a phase in which the $z$ fields are ordered and the gauge
modes are in the inverted XY high-temperature phase; for large $J$ and
small $\kappa$ (mixed molecular phase) gauge interactions are
long-ranged, while the spin degrees of freedom condense. Presumably,
for any $N$ the three transition lines meet at a multicritical point
at ($\kappa_{mc}, J_{mc})$, see Fig.~\ref{phdiasketch}.  This phase
diagram was proposed for $N=2$ in Refs.~\cite{MV-08,KMPST-08}, but it
should hold for any $N\ge 2$.

For any $N$, we expect the MH transition line to be continuous, in the
XY universality class, for any $J > J^*$, where $J^*$ may coincide
with the position of the multicritical point, i.e., $J^*\ge J_{mc}$.
Along this transition line the spins are expected to be frozen. They
should only act as spectators.

The transitions along the CM line are expected to have the same nature
as the $\kappa=0$ transition, at least for $\kappa < \kappa^*$, where
$\kappa^*$ must satisfy $\kappa^* \le \kappa_{mc}$.  As we said, we do
not expect the addition of $H_g$ for small $\kappa$ to change the
nature of the transition. Therefore, as it occurs for $\kappa = 0$, we
expect gauge modes to be irrelevant. This suggests that these
transitions are controlled by the LGW $\Phi^4$ field theory
(\ref{hlg}), in which gauge modes are effectively integrated out.
Thus, they should belong to the O(3) vector universality class for
$N=2$ and be of first order for $N\ge 3$.  Like the transition of the
CP$^{N-1}$ model for $\kappa=0$, the CM transition line is
characterized by the condensation of the gauge-invariant bilinear
operator (\ref{qdef}).

Finally, along the CH transition line, both scalar and gauge fields
change their long-distance behavior.  Therefore, we expect this
transition line to be described by the continuum AH model
(\ref{abhim}), whose RG flow predicts that continuous transitions may
be observed only for $N > N_c$, see Sec.~\ref{epsexp}.  Along the CH
line the gauge-invariant bilinear operator (\ref{qdef}) is expected to
be an appropriate order parameter.

\subsection{Nature of the multicritical point for $N=2$}
\label{muti}

As discussed above, the phase diagram of the noncompact lattice AH
model is characterized by three transition lines meeting at a
multicritical point. To discuss its nature within the field-theory
framework, it is crucial to identify the relevant critical modes.  For
$N=2$, we expect continuous transitions along the CM line with an O(3)
scalar order parameter, first-order transitions along the CH line, and
XY behavior along the MH line.  Thus, the nature of the multicritical
point is determined by the dynamics of two effective order-parameter
fields, a three-component and a two-component scalar field, which are
associated with the two continuous transition lines.  This hypothesis
seems quite reasonable, as the two transitions are associated with
different degrees of freedom: along the CM line the spin degrees of
freedom associated with the bilinear $Q_{\bm x}^{ab}$ defined in
Eq.(\ref{qdef}) condense, while gauge degrees of freedom control the
transitions along the CM line.

Note that this description is not expected to be appropriate for
$N\not=2$, and, in particular, for large values of $N$. In that case,
the CH transition line should be associated with the AH FP point, the
CM line is of first order, while the MH transitions belong to the XY
universality class.  In this case a correct description of the
multicritical point should describe the interaction of the XY order
parameter with the AH fields.  Thus, one should consider an extension
of the AH model that includes an additional two-component order
parameter.

For $N=2$, assuming that the multicritical behavior occurs from the
competition of two different scalar fields, we may investigate it
within a LGW framework.  If $\phi^a$ and $\varphi^a$ are a
three-component and a two-component field, respectively, the most
general $\Phi^4$ theory that is invariant under O(3)$\oplus$O(2)
transformations~\cite{LF-72,FN-74,NKF-74,CPV-03} is
\begin{eqnarray}
{\cal H} &=& 
\case{1}{2} \Bigl[ ( \partial_\mu \phi)^2  + (
\partial_\mu \varphi)^2\Bigr] + 
\case{1}{2} \Bigl( r_\phi \phi^2  + r_\varphi \varphi^2 \Bigr)
\nonumber\\
&&+
\case{1}{4!} \Bigl[ u_\phi (\phi^2)^2 + u_\varphi (\varphi^2)^2 + 
                           2 w \phi^2\varphi^2 \Bigr] \,,
\label{bicrHH} 
\end{eqnarray}
where $\phi^2\equiv \sum_{a=1}^3 \phi_a^2$ and $\varphi^2\equiv
\sum_{a=1}^2 \varphi_a^2$. Such model appears in several other
contexts, for instance, it is relevant for the behavior of high-$T$
superconductors (see Ref.~\cite{HPV-05} and references therein).  It
has been extensively studied in
Refs.~\cite{LF-72,FN-74,NKF-74,CPV-03,HPV-05}

\begin{figure}[tbp]
\includegraphics*[width=0.48\columnwidth]{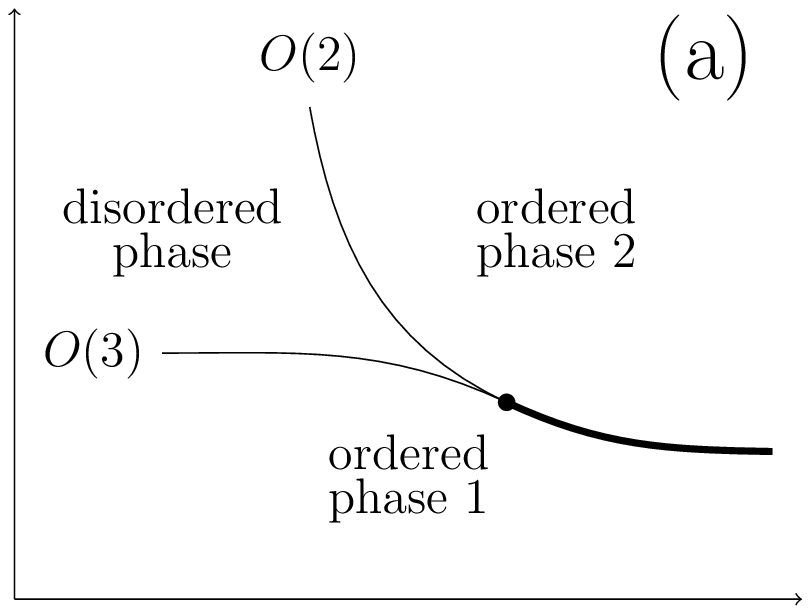}\\ 
\includegraphics*[width=0.48\columnwidth]{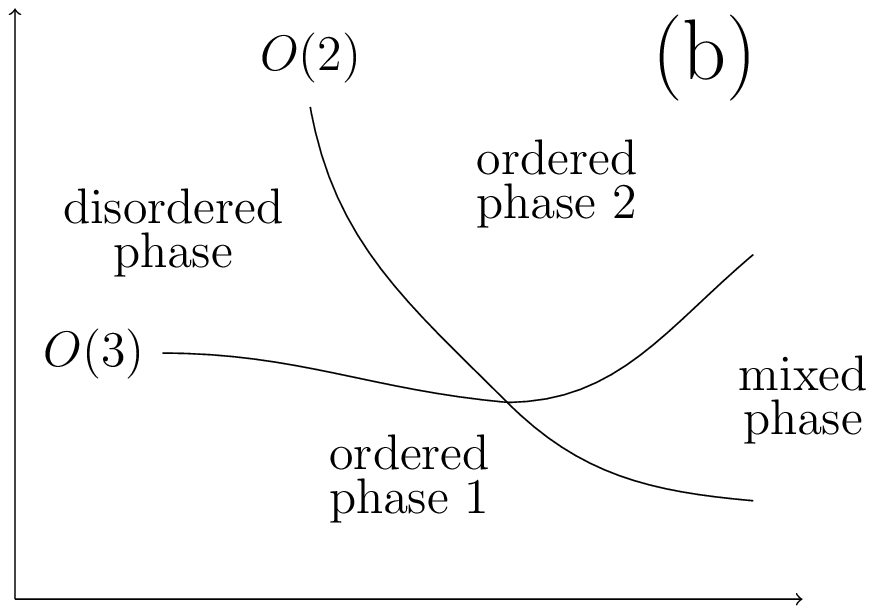}  
\includegraphics*[width=0.48\columnwidth]{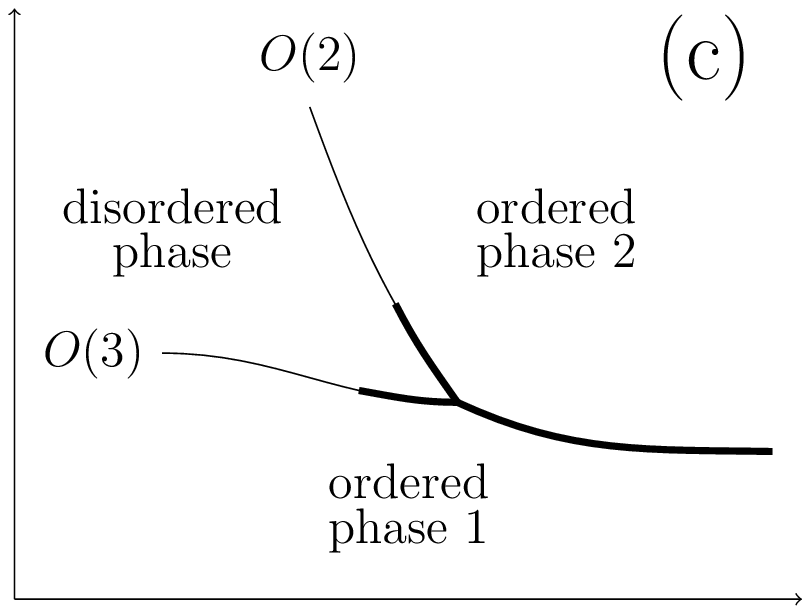}
  \caption{Sketches of the possible phase diagrams close to a
    multicritical point in the plane of the two relevant
    variables. Thin lines represent continuous phase transitions,
    while thick lines represent first-order transitions.  (a) Phase
    diagram with a bicritical point.  (b) Phase diagram with a
    tetracritical point.  (c) Phase diagram with a first-order
    bicritical point.}
\label{multicritfig}
\end{figure}

In the mean-field approximation~\cite{LF-72,FN-74,NKF-74}, two
possible phase diagrams are possible: a phase diagram, where two
continuous transition lines meet a first-order transition line, see
Fig.~\ref{multicritfig}(a)---the corresponding multicritical point is
called bicritical---and a phase diagram where four transition lines
meet, see Fig.~\ref{multicritfig}(b)---in this case the multicritical
point is called tetracritical. In our case, the multicritical point is
bicritical, the first-order transition line being identified with the
CH transition line.  Indeed, the numerical results reported in
Refs.~\cite{MV-08,KMPST-08} confirm that the transitions along the CH
line are of first order, at least for $\kappa$ not too large (they
apparently disagree only far from the multicritical point). To clarify
whether the bicritical transition is continuous or of first order, it
is necessary to analyze the RG flow of the multicritical $\Phi^4$
theory (\ref{bicrHH}): the multicritical transition can be continuous
only if a stable fixed point can be associated with the bicritical
point.

The analysis reported in Refs.~\cite{CPV-03,HPV-05} indicates that the
only stable fixed point is the decoupled fixed point describing
decoupled O(3) and O(2) critical behaviors, which is naturally
associated with a tetracritical point, see Fig.~\ref{multicritfig}(b).
There is no stable fixed point that can be associated with a
bicritical point. Indeed, fixed points that can be associated with a
bicritical behavior, for instance, the O(5) fixed point (in this case
there would be an enlargement of the symmetry at the multicritical
point) and the so-called biconical fixed point, are both
unstable~\cite{CPV-03,HPV-05}. In the absence of stable fixed points
that may be associated with a bicritical point, close to the
multicritical point the transitions are expected to be of first order
along all three lines, see Fig.~\ref{multicritfig}(c). Therefore, the
continuous O(3) and XY transition lines starting from the $\kappa=0$
and $J=\infty$ lines are expected to turn into first-order transition
lines before reaching the multicritical point.

The above LGW RG analysis predicts the CH transition line to be of
first order close to the multicritical point.  Increasing $\kappa$
along this line, the first-order transition should become weaker and
weaker (the latent decreases) as the O(4) continuous transition at
$\kappa=\infty$ is approached, with substantial crossover phenomena
occurring for large values of $\kappa$.  Alternatively, the
first-order transition could turn into a continuous transition line
belonging to a different universality class already for finite values
of $\kappa$.  We note that the existence of a corresponding
universality class is still controversial, see, e.g.,
Refs.~\cite{MV-04, Sandvik-07, MK-08, MV-08, KMPST-08, Sandvik-10,
  NCSOS-15, WNMXS-17, PV-20-mfcp, SZ-20}.

\section{Numerical results for the noncompact AH models}
\label{numres}

We now present a Monte Carlo (MC) study of the phase diagram of the
noncompact lattice AH model, for $N=2,4,10,15$ and $N=25$.  We perform
standard Metropolis updates of $A_{{\bm x},\mu}$ and ${\bm z}_{\bm
  x}$, combined with microcanonical updates of ${\bm z}_{\bm x}$ to
reduce autocorrelations, see, e.g., Ref.~\cite{PV-19-AH3d}.

As already noted, due to the peculiarities of the noncompact
formulation, we cannot consider finite systems with periodic boundary
conditions. Indeed, in this case there are gauge-invariant zero modes,
that make the model always ill-defined.  The zero modes correspond to
the so-called noncompact Polyakov lines, the gauge invariant sum of
the fields $A_{{\bm x},\mu}$ along nontrivial paths winding around the
lattice. Therefore, even if a maximal gauge fixing is added, the
partition function is still infinite. Under these conditions, it is
not clear whether finite size scaling (FSS) methods can be safely used
to investigate the critical behavior of the model.  To overcome this
problem, we adopt $C^*$ boundary conditions \cite{KW-91, LPRT-16}.
They preserve gauge invariance, providing a rigorous definition of the
partition function in a finite volume. Moreover, they essentially
preserve translational invariance. A detailed description is provided
in the App.~\ref{appcstar}.

\subsection{Observables and finite-size scaling analyses}
\label{obsfss}

We compute the energy density and the specific heat, defined as
\begin{eqnarray}
E = {1\over V} \langle H \rangle,\qquad
C ={1\over  V}
\left( \langle H^2 \rangle 
- \langle H \rangle^2\right),
\label{ecvdef}
\end{eqnarray}
where $V=L^3$.  We consider the two-point correlation function of the
gauge-invariant operator $Q_{\bm x}^{ab}$ defined in Eq.~(\ref{qdef}), 
\begin{equation}
G({\bm x}-{\bm y}) = \langle {\rm Tr}\, Q_{\bm x}Q_{\bm y} \rangle,  
\label{gxyp}
\end{equation}
where the translation invariance of the system has been taken into
account (note that $Q_{\bm x}$ is periodic also in the presence of
$C^*$ boundary conditions, see App.~\ref{appcstar}). The
susceptibility and the (second-moment) correlation length are defined
by the relations
\begin{eqnarray}
&&\chi =  \sum_{{\bm x}} G({\bm x}) = 
\widetilde{G}({\bm 0}), 
\label{chisusc}\\
&&\xi^2 \equiv  {1\over 4 \sin^2 (\pi/L)}
{\widetilde{G}({\bm 0}) - \widetilde{G}({\bm p}_m)\over 
\widetilde{G}({\bm p}_m)},
\label{xidefpb}
\end{eqnarray}
where $\widetilde{G}({\bm p})=\sum_{{\bm x}} e^{i{\bm p}\cdot {\bm x}}
G({\bm x})$ is the Fourier transform of $G({\bm x})$, and ${\bm p}_m =
(2\pi/L,0,0)$.

It is convenient to introduce RG-invariant quantities, such as the
Binder parameter
\begin{equation}
U = \frac{\langle \mu_2^2\rangle}{\langle \mu_2 \rangle^2} \,, \qquad
\mu_2 = \frac{1}{V^2}  
\sum_{{\bm x},{\bm y}} {\rm Tr}\,Q_{\bm x} Q_{\bm y}\,,
\label{binderdef}
\end{equation}
and 
\begin{equation}\label{rxidef}
R_{\xi}=\xi/L\,.
\end{equation}
At continuous phase transitions they are expected to scale as
\cite{PV-02} (we denote by $R$ a generic RG invariant quantity)
\begin{eqnarray}
R(\beta,L) &\approx & f_R(X) \,,\quad X =(\beta-\beta_c)\,L^{1/\nu}\,,
 \label{scalbeh}
\end{eqnarray}
where $\nu$ is the critical exponent associated with the correlation
length, and $\beta$ is the parameter we vary in the system (in the
following sections $J$ will play the role of $\beta$). Scaling
corrections decaying as $L^{-\omega}$ have been neglected in
Eq.~(\ref{scalbeh}), where $\omega$ is the exponent associated with
the leading irrelevant operator. The function $f_R(X)$ is universal up
to a multiplicative rescaling of its argument.  In particular,
$U^*\equiv f_U(0)$ and $R_\xi^*\equiv f_{R_\xi}(0)$ are universal,
depending only on the boundary conditions and the aspect ratio of the
lattice.  Since $R_\xi$ defined in Eq.~\eqref{rxidef} is an increasing
function of $\beta$, we can write
\begin{equation}\label{uvsrxi}
U(\beta,L) = F_U(R_\xi) + O(L^{-\omega})\,,
\end{equation}
where $F_U$ now depends on the universality class, boundary conditions
and lattice shape, without any nonuniversal multiplicative factor.
The scaling relation \eqref{uvsrxi}, which does not involve any
nonuniversal parameter, is particularly convenient to test
universality-class predictions and to identify weak first-order
transitions~\cite{PV-19-CP,PV-19-AH3d}.

At first-order transitions the probability distributions of the energy
and of the magnetization are expected to show a double peak for large
values of $L$.  However, in order to definitely identify a first-order
transition, it is necessary to perform a more careful analysis of the
large-$L$ scaling behavior of the distributions or, equivalently, of
the specific heat and of the Binder cumulants
~\cite{CLB-86,VRSB-93,LK-91,CPPV-04,CNPV-14}.  Regarding the specific
heat $C$, for each lattice size $L$, there exists a value $\beta_{{\rm
    max},C}(L)$ of $\beta$ where $C$ takes its maximum value $C_{\rm
  max}(L)$. For large volumes, we have \cite{CLB-86}
\begin{eqnarray}
&&C_{\rm max}(L) = \frac{V}{4} \Delta_h^2 + O(1),
\label{cmaxsc}\\
&&\beta_{{\rm max},C}(L)-\beta_c\approx c\,V^{-1} \label{betamax},
\end{eqnarray}
where $\Delta_h = E_+ - E_-$ is the latent heat, and 
$E_+$ and $E_-$ are the values of the energy corresponding to the
two maxima of the energy-density distribution.

As discussed in Ref.~\cite{PV-19-CP,PV-19-AH3d} the Binder parameter
is a particularly convenient quantity to identify first-order
transitions. In this case~\cite{VRSB-93}, for each $L$, $U(\beta,L)$
has a maximum $U_{\rm max}(L)$ at $\beta = \beta_{{\rm max},U}(L) <
\beta_c$ which scales, for sufficiently large values of $L$, as
\begin{eqnarray}
&&U_{\rm max} \sim a\,V + O(1)\,,\label{umaxsca}\\ &&\beta_{{\rm
    max},U}(L) - \beta_c \approx b \,V^{-1}\,.
\label{bpeakU}
\end{eqnarray}
This should be contrasted with the behavior at a continuous
transition, where $U$ is always finite. Thus, we can distinguish
first-order from continuous transitions by looking at the behavior of
$U$ as $L$ increases. In particular, the absence of scaling when
plotting the data of $U$ versus $R_\xi$ may be considered as an
evidence in favor of a first-order transition.

Note that FSS also holds at first-order
transitions~\cite{NN-75,FB-82,PF-83,CPPV-04}, although it is more
sensitive to the geometry and to the nature of the boundary
conditions~\cite{CNPV-14}; for instance, FSS differs for boundary
conditions that favor or do not favor the different phases coexisting
at the transition~\cite{HPV-18,PRV-18}. In the case of 3D cubic
systems with boundary conditions respecting translation invariance,
such as periodic boundary conditions, FSS is typically characterized
by an effective exponent $\nu=1/d=1/3$, so that
$\alpha/\nu=d=3$. Thus, effective exponents that decrease towards 1/3
as $L$ increases, signal a discontinuous transition.

\subsection{The noncompact lattice AH model at $N=2$}
\label{n2res}

\begin{figure}[t]
  \includegraphics*[width=0.8\columnwidth]{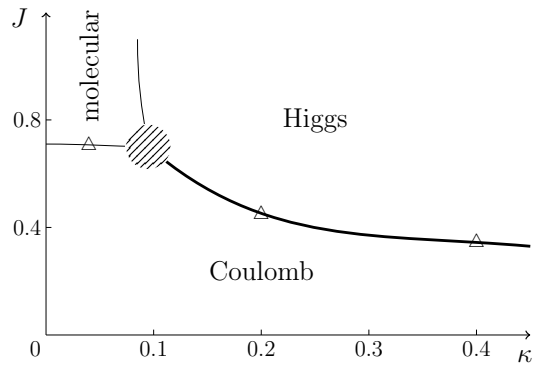}
  \caption{Phase diagram of the noncompact $N=2$ lattice AH model, as
    obtained from the data of Refs.~\cite{MV-08,KMPST-08}. Triangles
    correspond to the transition points obtained in the present work.
    The shadowed blob indicates the region where the transition lines
    meet and the transitions along the three lines are predicted to be
    of first order, see Sec.~\ref{muti}.}
\label{N2phadia}
\end{figure}

\begin{figure}[tbp]
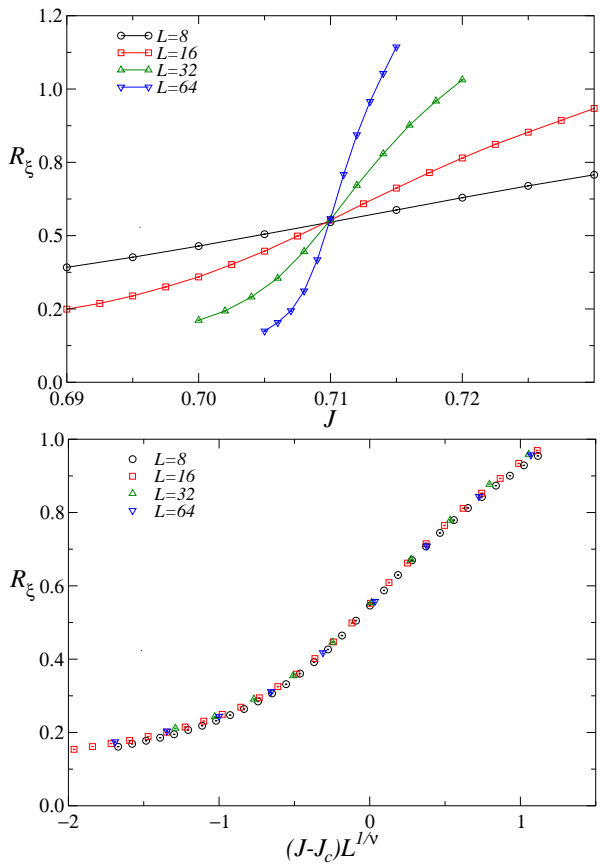

\includegraphics*[scale=\graphicscale]{rxiJ_nccp1_kappa0p04.eps}
\includegraphics*[scale=\graphicscale]{rxiJ_nccp1_kappa0p04_scaled.eps}
\caption{Plot of $R_\xi$ for $N=2$ and $\kappa=0.04$ 
(along the CM transition line) 
for lattice sizes
  up to $L=64$.  Top panel: $R_{\xi}$ versus $J$; data show a crossing
  point at $J=J_c = 0.7099(1)$.  Bottom panel: $R_{\xi}$ versus
  $(J-J_c)L^{1/\nu}$ using the O(3) critical exponent
  $\nu=0.7117$~\cite{Hasenbusch-20,KP-17,HV-11,CHPRV-02}. }
\label{N2O3rxi}
\end{figure}

\begin{figure}[tbp]
\includegraphics*[scale=\graphicscale]{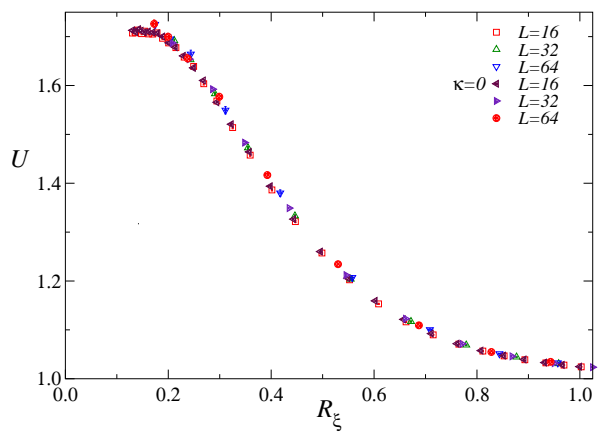}
\caption{Plot of $U$ versus $R_\xi$ for the noncompact $N=2$ lattice
  AH model at $\kappa=0.04$ and $\kappa = 0$ (along the CM transition
  line). All data fall onto a single curve, providing a robust
  evidence that the transition belongs to the O(3) universality class.
}
\label{uvsrxin2o3}
\end{figure}

\begin{figure}[tbp]
\includegraphics*[scale=\graphicscale]{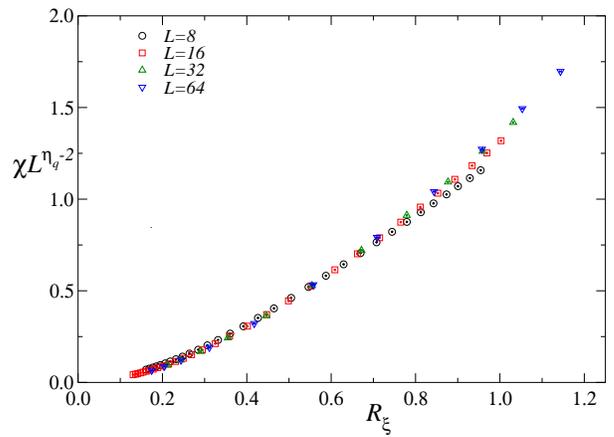}
\caption{Plot of $\chi/L^{2-\eta_q}$ at $\kappa=0.04$ versus $R_{\xi}$
  along the CM transition line, using the O(3) value
  $\eta_q=0.0378$~\cite{Hasenbusch-20,KP-17,HV-11,CHPRV-02}.  Results
  for the $N=2$ AH model.  }
\label{chin2o3}
\end{figure}

\begin{figure}[tbp]
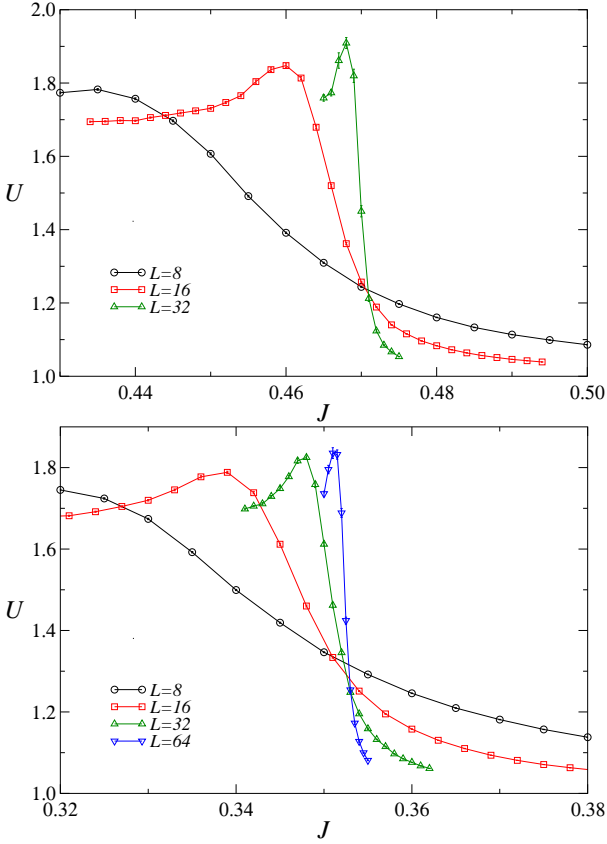

\includegraphics*[scale=\graphicscale]{uJ_nccp1_kappa0p2.eps}
\includegraphics*[scale=\graphicscale]{uJ_nccp1_kappa0p4.eps}
\caption{Binder parameter $U$ for the $N=2$ lattice AH model, at
  $\kappa=0.2$ and for $L \le 32$ (top), and at $\kappa=0.4$ for $L
  \le 64$ (bottom).  Both transitions should be along the CH
  transition line.  }
\label{ujn2}
\end{figure}

\begin{figure}[tbp]
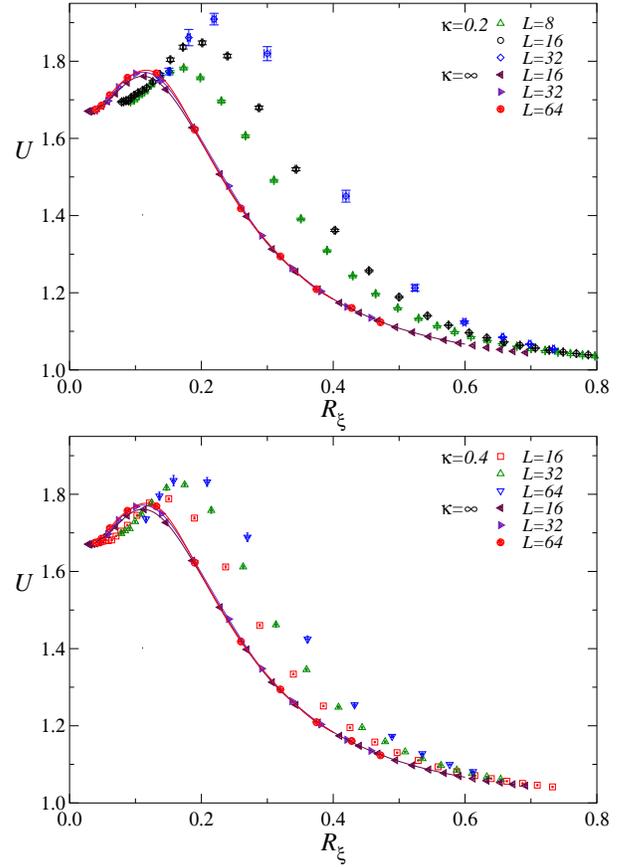

\includegraphics*[scale=\graphicscale]{urxi_nccp1_kappa0p2.eps}
\includegraphics*[scale=\graphicscale]{urxi_nccp1_kappa0p4.eps}
\caption{Estimates of $U$ versus $R_\xi$ for the $N=2$ AH model at
  $\kappa=0.2$ (top) and $\kappa= 0.4$ (bottom), thus along the CH
  transition line.  For comparison $O(4)$ data, corresponding to
  $\kappa=\infty$, are also reported; the continuous lines are cubic
  spline interpolations and have the only purpose of guiding the
  eye. }
\label{uvsrxin2lk}
\end{figure}

The lattice AH model with $N=2$ has already been studied in
Refs.~\cite{MV-08,KMPST-08}, obtaining the phase diagram shown in
Fig.~\ref{N2phadia}.  We present a different FSS analysis, using $C^*$
boundary conditions, which is not affected by the pathologies of
periodic boundary conditions, see App.~\ref{appcstar}.  In particular,
we present results along the CM and CH transition lines, see
Fig.~\ref{N2phadia}.

To begin with, we discuss the critical behavior along the CM line
starting at $\kappa=0$, $J =J_c= 0.7102(1)$.  We present a FSS
analysis of MC data taken at fixed $\kappa=0.04$. The data of $R_\xi$,
see Fig.~\ref{N2O3rxi}, show a crossing point at $J_c = 0.7099(1)$,
which is very close to the critical value at $\kappa=0$. The plot of
the Binder parameter $U$ versus $R_\xi$, see Fig.~\ref{uvsrxin2o3},
shows that the critical behavior is the same for $\kappa = 0$ and
$\kappa = 0.04$. Since the CP$^1$ transition belongs to the O(3)
vector universality class, the same is expected for the transition at
$\kappa = 0.04$. The O(3) scaling is also confirmed by the scaling of
$R_{\xi}$ and of the susceptibility: this is observed when plotting
$R_{\xi}$ versus $(J-J_c)L^{1/\nu}$ and the ratio $\chi/L^{2-\eta_q}$
versus $R_\xi$ using the O(3) critical exponents $\nu\approx 0.7117$
and $\eta_q\approx 0.0378$~\cite{Hasenbusch-20, KP-17, HV-11,
  CHPRV-02}, see Figs.~\ref{N2O3rxi} and \ref{chin2o3}.

To investigate the nature of the transitions along the CH line, we
have performed MC simulations at $\kappa=0.2$ and $\kappa=0.4$. As we
shall see, in both cases the data clearly favor a first-order
transition. This confirms the analysis reported in
Ref.~\cite{KMPST-08}. It disagrees instead with Ref.~\cite{MV-08},
that claimed the transition for $\kappa=0.4$ to be continuous.
Fig.~\ref{ujn2} reports the behavior of the Binder parameter $U$ for
$\kappa = 0.2$ and 0.4: data indicate the presence of a transition at
$J=J_c\approx 0.472$ for $\kappa = 0.2$ and at $J = J_c\approx 0.353$
for $\kappa=0.4$, the latter result being in agreement with
Ref.~\cite{MV-08}. In Fig.~\ref{uvsrxin2lk} we plot $U$ versus $R_\xi$
for both $\kappa = 0.2$ and 0.4. Data do not scale, providing evidence
in favor of a weak first-order transition for both $\kappa = 0.2$ and
0.4 (see the discussion in Sec.~\ref{obsfss}).  Comparing the data, we
observe that the first-order transition becomes weaker with increasing
$\kappa$, in agreement with the expectation that the latent heat
vanishes in the limit $\kappa\to\infty$.  For comparison we also
report data for the O(4) vector model with the $C^*$ boundary
conditions, to identify possible crossover effects, which indeed seem
to appear.

\subsection{The noncompact lattice AH model at large $N$}
\label{nlargeres}

\begin{figure}[b]
\includegraphics*[scale=\graphicscale]{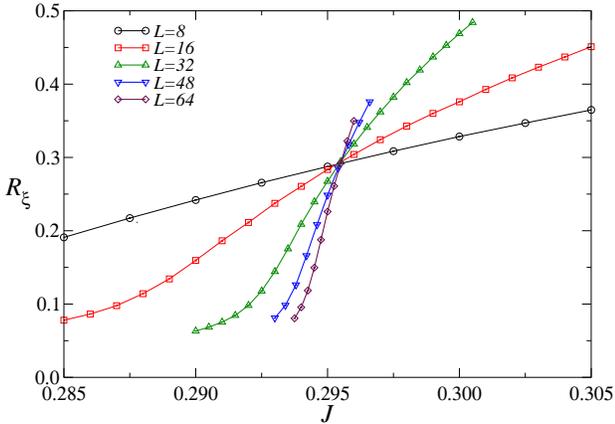}
\caption{Estimates of $R_{\xi}\equiv\xi/L$ for the $N=25$ lattice AH
  model at $\kappa=0.4$, for several lattice sizes up to $L=64$.  
  The data show clearly a crossing point providing an estimate of the
  critical value $J_c\approx 0.2955$.}
\label{rxin25lk}
\end{figure}

\begin{figure}[tbp]
\includegraphics*[scale=\graphicscale]{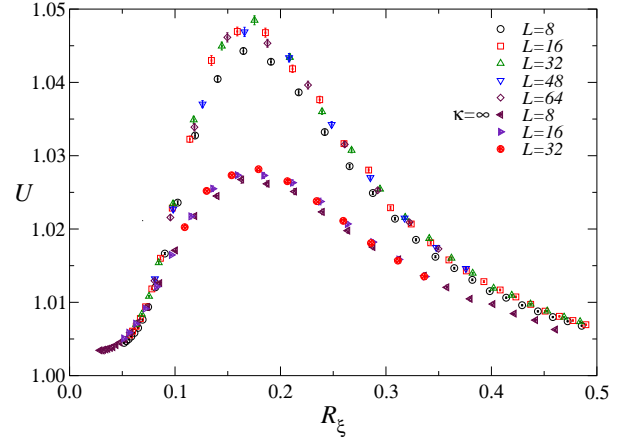}
\caption{Plot of $U$ versus $R_\xi$ for the $N=25$ lattice AH model at
  $\kappa=0.4$. For comparison we also report data for the O(50)
  vector model, corresponding to the $\kappa\to\infty$ limit.  }
\label{uvsrxin25lk}
\end{figure}

We now present numerical results for large values of $N$.  As
discussed in Sec.~\ref{phadia}, we expect a phase diagram
characterized by three phases also for $N>2$. However, for $N\not=2$,
the CM transition line is expected to be of first order, as for
$\kappa = 0$. On the other hand, continuous transitions may appear
along the CH transition line for sufficiently large $N$, since the RG
flow of the continuum AH theory has a stable fixed point, see
Sec.~\ref{models}.

\begin{figure}[tbp]
\includegraphics*[scale=\graphicscale]{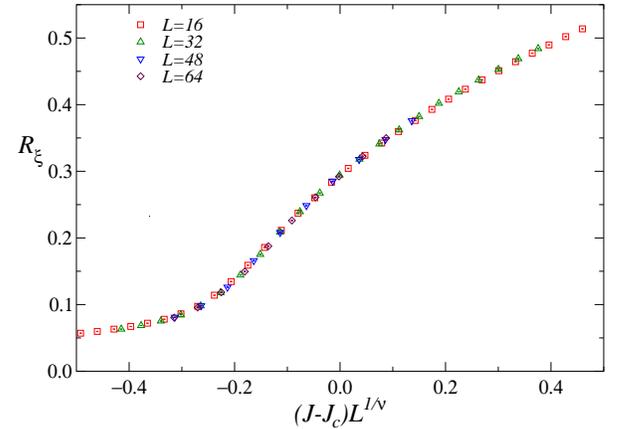}
\caption{Plot of $R_{\xi}\equiv\xi/L$ versus $(J-J_c) L^{1/\nu}$ 
  at at $\kappa=0.4$ for $N=25$. We use
  $J_c=0.295511$ and $\nu=0.802$.}
\label{rxiscan25lk}
\end{figure}

We first consider the $N=25$ model, performing simulations for
$\kappa=0.4$.  The data of $R_\xi$, see Fig.~\ref{rxin25lk}, show a
crossing point for $J_c\approx 0.295$, that we identify as a
transition point along the CH line.  To understand the order of the
transition, we plot the Binder parameter $U$ versus $R_\xi$, see
Fig.~\ref{uvsrxin25lk}.  Data scale nicely, strongly suggesting that
the transition is continuous.  To determine the critical exponent
$\nu$ and obtain a more accurate estimate of the critical point, we
fit $U$ and $R_\xi$ to
\begin{equation}
R(J,L) = f_R(X)\,, \qquad X = (J-J_c) \,L^{1/\nu}\, ,
\label{fitR-senzacorrezioni}
\end{equation}
using a polynomial approximation for $f_R(X)$. To estimate the role of
the scaling corrections we restrict the fit to the data satisfying
$L\ge L_{\rm min}$, with $L_{\rm min} = 16, 32$. For $L_{\rm min}=16$
we obtain $\nu = 0.789(2)$, $\nu = 0.785(1)$ from the analysis of $U$
and $R_\xi$, respectively.  For $L_{\rm min} = 32$, we find $\nu =
0.782(5), 0.796(2)$. The variation of the results appears larger than
the statistical errors, indicating that scaling corrections are
significant, at the level of precision of our data. We have thus
performed fits that include scaling corrections, fitting the data to
\begin{equation}
R(J,L) = f_R(X) + L^{-\omega} g_R(X) \,,
\label{Rfit-corrections}
\end{equation}
using a polynomial approximation for both scaling functions. To
improve the accuracy of the estimates, we have performed a combined
fit of the two observables. It turns out that our data are not precise
enough to allow us to determine the exponent $\omega$. The $\chi^2$ of
the fit takes essentially the same value for any $\omega \gtrsim
0.8$. Correspondingly, $\nu$ varies between 0.796 and 0.808, with a
statistical error of 0.002 at fixed $\omega$. This analysis allows us
to estimate $\nu = 0.802(8)$. The quality of the fit is excellent, as
can be seen from Fig.~\ref{rxiscan25lk}.  The estimate of $\nu$ is
very different from that corresponding to the O(50) vector model,
$\nu\approx 0.96$ (we use here the large-$N$ expansion of $\nu$),
confirming the instability of the O(50) fixed point in the AH field
theory. This is also confirmed by the comparison of the plots of $U$
versus $R_\xi$, see Fig.~\ref{uvsrxin25lk}, where we report data for
the O(50) vector model, i.e., $U$ and $R_\xi$ as obtained from the
spin-2 correlation function in the vector model.

The estimate of $\nu$ is very close to the estimate obtained using the
$1/N$ expansion at order $1/N$~\cite{HLM-74,YKK-96,MZ-03,KS-08},
$\nu_{ln} = 0.805$, see Eq.~(\ref{nulargen}). The large-$N$ expansion
appears to be very accurate for $\nu$ at $N=25$. It is interesting to
note that the exponent $\omega$ is equal to 1 for $N=\infty$ and thus
we expect it to be close to 1 also for $N=25$. Using this information,
we can verify that the large-$N$ expansion is probably accurate for
the exponent $\nu$ at the level of a few per mille.  Indeed, our
combined fits give $\nu = 0.805(2)$ and $\nu = 0.808(2)$ for $\omega =
1$ and 0.8, respectively.

\begin{figure}[tbp]
\includegraphics*[scale=\graphicscale]{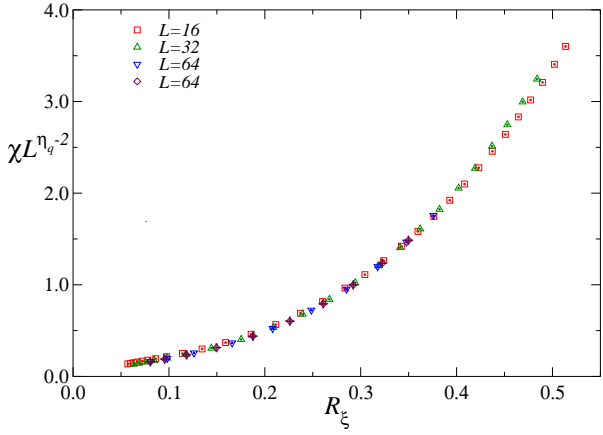}
\caption{Plot of $\chi L^{\eta_q - 2}$ versus $R_\xi$ 
  for the $N=25$ lattice AH model at
  $\kappa=0.4$.  We use $\eta_q = 0.883$.
  }
\label{chin25lk}
\end{figure}

The analysis also provides estimates of $J_c$ and of the universal
quantities $R^*_\xi$ and $U^*$, the last two quantities being the
asymptotic values ($L\to \infty$) of $R_\xi$ and $U$ computed for $J =
J_c$. The results are reported in Table~\ref{tabln}.  We have also
analyzed the susceptibility $\chi$ in order to determine the exponent
$\eta_q$. We have performed fits to $\chi = L^{2-\eta_q}
f_\chi(R_\xi)$, which has the advantage that neither $J_c$ nor $\nu$
appear in the fitting function.  We obtain $\eta_q = 0.923(1)$ and
$\eta_q = 0.901(1)$, if we only consider the data with $L \ge 16$ and
32, respectively.  There are clearly scaling corrections. We have
therefore performed fits to
\begin{equation}
   \chi = L^{2-\eta_q} \left[
     f_\chi(R_\xi) + L^{-\omega} g_\chi(R_\xi) \right],
\label{chifit}
\end{equation}
where we use polynomial approximations for $f_\chi(R_\xi)$ and
$g_\chi(R_\xi)$.  The $\chi^2$ of the fit has a shallow minimum for
$0.9\lesssim \omega \lesssim 1.5$, As $\omega$ varies in this
interval, $\eta_q$ varies from 0.880(5) and 0.887(3). We thus end up
with the estimate $\eta_q = 0.883(7)$.  The corresponding scaling plot
is shown in Fig.~\ref{chin25lk}: data scale quite precisely onto a
single curve for $L\ge 32$. Note that the final estimate is
essentially consistent with the large-$N$ result $\eta_{q,ln} =
0.870$, see Eq.~(\ref{etalargen}).

\begin{table*}
\begin{tabular}{rccccccc}
\hline\hline 
$N$ &  $J_c$ &  $\nu$ & $\nu_{ln}$ &
$\eta_q$ & $\eta_{q,ln}$  & $R_\xi^*$ & $U^*$ \\
\hline
25  &  0.295511(4) & 0.802(8) & 0.805 & 0.883(7) & 0.870 & 0.29405(5) & 
   1.0254(1) \\
15  &  0.309798(6) & 0.721(3) & 0.676 & 0.815(10)& 0.784 & 0.316(1) & 
   1.0433(3) \\
10 &   0.32187(3)  & 0.64(2)  & 0.514 & 0.74(2)  & 0.678 & 0.341(8) &
   1.0621(4) \\
\hline\hline
\end{tabular}
\caption{We report the estimates of the critical coupling $J_c$, of
  the critical exponents $\nu$ and $\eta_q$, and of the universal
  critical values $R_\xi^*$ and $U^*$ for $C^*$ boundary conditions at
  the continuous transitions observed for $N=10, 15$, and $N=25$ along
  the CH transition line ($\kappa=0.4$).  We also report the estimates
  $\nu_{\rm ln}$ and $\eta_{q,{\rm ln}}$ obtained from the $O(1/N)$
  approximations reported in Eqs.~(\ref{nulargen}) ad
  (\ref{etalargen}).}
  \label{tabln}
\end{table*}
\begin{figure}[tbp]
\includegraphics*[scale=\graphicscale]{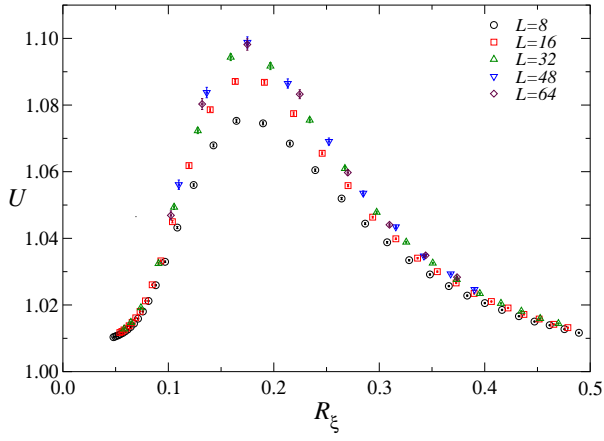}
\caption{Plot of $U$ versus $R_\xi$ for the $N=15$ lattice AH model at
  $\kappa=0.4$. 
  }
\label{uvsrxin15lk}
\end{figure}

\begin{figure}[tbp]
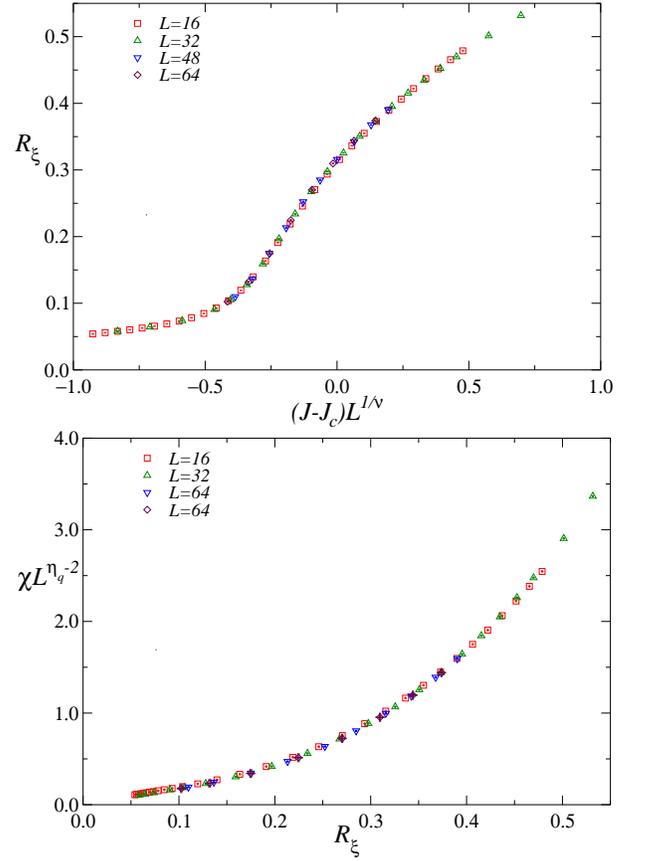

\includegraphics*[scale=\graphicscale]{rxiJ_nccp14_kappa0p4_scaled.eps}
\includegraphics*[scale=\graphicscale]{chirxi_nccp14_kappa0p4_scaled.eps}
\caption{Top panel: $R_{\xi}\equiv\xi/L$ versus $(J-J_c) L^{1/\nu}$
  for $J_c=0.309798$ and $\nu = 0.721$.  Bottom panel: $L^{-2+\eta_q}
  \chi$ versus $R_{\xi}$ for $\eta_q = 0.815$. Results for the $N=15$
  lattice AH model at $\kappa=0.4$ for several lattice sizes up to
  $L=64$.  }
\label{rxin15lk}
\end{figure}

We have also observed a transition for $N=15$ along the line $\kappa =
0.4$.  Indeed, data for $R_{\xi}$ and $U$ show a crossing point for
$J_c\approx 0.31$.  To identify the order of the transition, we plot
$U$ as a function of $R_{\xi}$, see Fig.~\ref{uvsrxin15lk}.  Scaling
corrections are clearly visible, but note that the data for $L\ge 32$
lie on top of each other.  The Binder parameter does not increase with
the size, indicating that the transition is continuous. To estimate
the critical exponents, we have repeated the analysis we did for
$N=25$. Scaling corrections are significant, as it appears from
Fig.~\ref{uvsrxin15lk}. Therefore, they must be taken into account to
obtain reliable estimates. We perform combined fits of $R_\xi$ and $U$
to the Ansatz (\ref{Rfit-corrections}).  If we only include data with
$L\ge 16$, the $\chi^2$ of the fit is essentially constant for
$1\lesssim \omega \lesssim 2$. In this range of values of $\omega$,
the exponent $\nu$ varies between 0.722(2) and 0.720(2), allowing us
to obtain the final estimate 0.721(3).  Results for $J_c$ and for the
critical values $R^*_\xi$ and $U$ are collected in Table~\ref{tabln}.
The corresponding scaling plot is reported in Fig.~\ref{rxin15lk}: the
scaling behavior is excellent.  We have also determined the exponent
$\eta_q$, fitting $\chi$ to Eq.~(\ref{chifit}). We obtain $\eta_q =
0.815(10)$.  Again we can compare the results for $\nu$ and $\eta_q$
with the large-$N$ results. In this case, see Table~\ref{tabln}, some
discrepancies are observed, indicating that for $N=15$ the corrections
of order $1/N^2$ are now significant.

\subsection{The noncompact lattice AH model for intermediate values of $N$}
\label{nmediumres}

\begin{figure}[tbp]
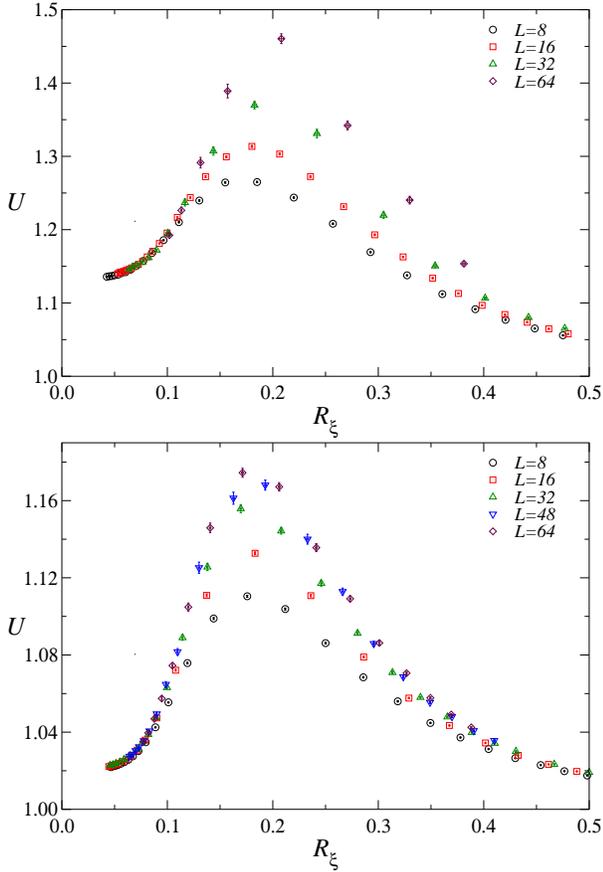

\includegraphics*[scale=\graphicscale]{urxi_nccp3_kappa0p4.eps}
\includegraphics*[scale=\graphicscale]{urxi_nccp9_kappa0p4.eps}
\caption{Top panel: estimates of $U$ versus $R_\xi$ for the $N=4$
  lattice AH model at $\kappa=0.4$.  Bottom panel: estimates of of $U$
  versus $R_\xi$ for the $N=10$ lattice AH model at $\kappa=0.4$.  }
\label{uvsrxin4and10lk}
\end{figure}

\begin{figure}[tb]
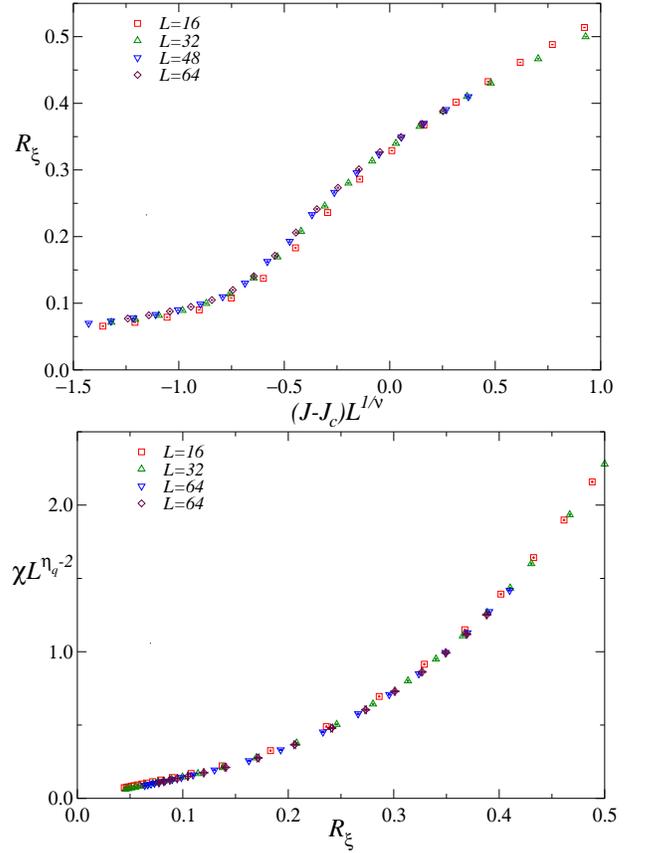

\includegraphics*[scale=\graphicscale]{rxiJ_nccp9_kappa0p4_scaled.eps}
\includegraphics*[scale=\graphicscale]{chirxi_nccp9_kappa0p4_scaled.eps}
\caption{Top: Plot of $R_{\xi}$ versus $(J - J_c) L^{1/\nu}$; we use
  $J_c = 0.32187$ and $\nu = 0.64$.  Bottom: Plot of $\chi L^{\eta_q -
    2}$ versus $R_\xi$; we use $\eta_q = 0.74$.  Results for the
  $N=10$ lattice AH model at $\kappa=0.4$, for several lattice sizes
  up to $L=64$.  }
\label{RUvsXn10}
\end{figure}

In the previous sections, we observed that the CH transition line is
of first order for $N=2$, while it is continuous for $N=15, 25$.
Therefore, there must be an intermediate number $N_\ell$, where the
nature of the transition changes: for $N \ge N_\ell$ the transition is
continuous, while for $N < N_\ell$ it is of first order.  An analogous
behavior is predicted by the continuum AH field theory, as discussed
in Sec.~\ref{epsexp}.  The results of the previous sections give
$2<N_\ell<15$.  In this section we present some results for $N=4$ and
$N=10$, that further constrain $N_\ell$.

In Fig.~\ref{uvsrxin4and10lk} we plot $U$ versus $R_{\xi}$ for $N=4$
along the line $\kappa=0.4$.  As it occurs for $N=2$, see
Sec.~\ref{n2res}, the data do not scale. Moreover, the Binder
parameter $U$ has a maximum that increases with $L$. The data
therefore favor a first-order transition, allowing us to conclude
$N_\ell>4$.  Note that the increase of the maximum of $U$ with the
size signals that the first-order transition is stronger than for
$N=2$. The transition is however too weak to allow us to reliably
estimate the latent heat using lattice sizes up to $L=64$. This is a
serious obstruction to what would be the natural strategy to estimate
$N_\ell$, i.e., to determine the behavior of the latent heat as a
function of $N$.

To obtain an upper bound on $N_\ell$, we performed simulations for
$N=10$ at $\kappa=0.4$.  Estimates of $U$ against $R_{\xi}$ are shown
in Fig.~\ref{uvsrxin4and10lk}. The maximum of the Binder parameter
appears to increase with the size for small values of $L$, but the
results for $L=64$ and $L=48$ apparently fall one on top of the
other. Therefore, data suggest that the transition is continuous,
implying the upper bound $N_\ell < 10$.

To determine the critical exponent $\nu$ for $N=10$, we perform a
combined fit of $U$ and $R_\xi$ to Eq.~(\ref{Rfit-corrections}). In
this case the fit is sensitive to $\omega$ and indeed the fit gives
$\omega = 1.05(10)$ and $\nu = 0.642(4)$.  However, the $\chi^2$ per
degree of freedom (DOF) is quite large, $\chi^2/\hbox{DOF} \approx
16$. Clearly, there are still significant scaling corrections that are
not taken into account by the scaling Ansatz
(\ref{Rfit-corrections}). Therefore, the statistical errors are not
reliable: systematic errors due to the neglected scaling corrections
are significantly larger. To get a rough idea of the size of the
systematic errors, we can compare the previous estimate of $\nu$ with
those obtained by using the simpler Ansatz
(\ref{fitR-senzacorrezioni}). For $R_\xi$, if we only include the data
with $L\ge 32$, we obtain $\nu = 0.658(1)$. This suggests that a
reliable estimate for $\nu$ might be $\nu = 0.64(2)$. Using the same
criterion for $J_c$, $R_\xi^*$, and $U^*$, we obtain the estimates
reported in Table~\ref{tabln}. The scaling plot of $R_\xi$ are
reported in Fig.~\ref{RUvsXn10}. Scaling deviations are clearly
visible for $L=16$. We have also determined the exponent $\eta_q$. The
susceptibility has been fitted to the Ansatz (\ref{chifit}). The
$\chi^2$ is little sensitive to $\omega$ and is essentially constant
for $\omega \gtrsim 1.4$. Correspondingly, $\eta_q$ varies between
0.745(4) and 0.759(2). However, note that we expect $\omega \approx 1$
and thus we have conservatively considered the larger interval $\omega
\gtrsim 0.8$.  Since $\eta_q = 0.727(6)$ for $\omega = 0.8$, we end up
with the final estimate $\eta_q = 0.74(2)$. The quality fo the fit is
excellent, see~Fig.~\ref{RUvsXn10}.

In conclusion, the numerical results for $N=4$ and $N=10$ allow us to
conclude that
\begin{equation}
4<N_\ell<10\,.
\label{nellb}
\end{equation}
A more precise determination of $N_\ell$ would require a substantially
bigger computational effort, so that we do not pursue this issue
further.

The upper bound on $N_\ell$ provides an upper bound on the number
$N_c$, the smaller value of $N$ for which the 3D AH field theory has a
stable fixed point, as discussed in Sec.~\ref{epsexp}.  If we assume
that the transitions observed for $N\ge 10$ can all be associated with
the field theory stable fixed point---we provided evidence for that in
the previous sections---we can conclude that
\begin{equation}
N_c<10\,.
\label{ncub}
\end{equation} 
On the other hand the lower bound on $N_{\ell}$ cannot be
straightforwardly extended to $N_c$. Indeed, the evidence of a
first-order transition for a lattice model does not exclude the
possibility that the corresponding field theory has a stable fixed
point, since the given lattice model might be outside its attraction
domain.  We finally note that the bound Eq.~(\ref{ncub}) is in
substantial agreement with the estimate $N_c =12.2(3.9)$ obtained by
the analysis of the four-loop $\varepsilon$ expansion~\cite{IZMHS-19},
mentioned in Sec.~\ref{epsexp}.

\section{Conclusions}
\label{conclu}

We have investigated the phase diagram and nature of the phase
transitions of the 3D multicomponent lattice AH model with noncompact
gauge fields. Our study confirms the existence of significant
differences with the lattice AH model with compact gauge fields, both
for small and large $N$~\cite{PV-19-AH3d}.  As sketched in
Fig.~\ref{phdiasketch}, the phase diagram of the noncompact model with
$N\ge 2$ is generally characterized by three phases: (i)~the Coulomb
phase, where the global SU($N$) symmetry is unbroken and the
electromagnetic correlations are long-ranged; (ii)~the Higgs phase,
where the local bilinear operator $Q_{\bm x}$ [cf. Eq.~(\ref{qdef})]
condenses, breaking the global SU($N$) symmetry, and electromagnetic
correlations are gapped; (iii)~a mixed molecular phase, where the
local bilinear $Q_{\bm x}$ condenses, but the electromagnetic field
remains ungapped.  We recall that the phase diagram of the compact
lattice AH model presents only two phases. They are characterized by
the condensation of the bilinear operator $Q_{\bm x}$, while gauge
fields are always in the confined phase~\cite{PV-19-AH3d}.

We have studied in detail the Coulomb-to-Higgs transition line that
ends at the O($2N$) transition point for $\kappa=\infty$ (i.e., for
vanishing gauge coupling), and the Coulomb-to-molecular transition
line that ends at the CP$^{N-1}$ transition. Transitions along the CH
line, if continuous, are expected to be associated with the stable
fixed point of the RG flow of the continuum AH model (\ref{abhim}).
On the other hand, transitions along the CM line should be described
by the LGW theory (\ref{hlg}), because gauge correlations do not play
a relevant role.

We summarize the behavior along the three transition lines as follows:

\noindent
(i) The CH transitions separating the Coulomb and Higgs phases are
weak first-order transitions for $N=2$. The same behavior is expected
for sufficiently small $N$. Indeed, a first-order transition is
observed for $N=4$. As $N$ increases, the transitions become
continuous, at least not too close to the multicritical point.  We
observe continuous transitions for $N\ge 10$. The corresponding
critical behavior turns out to belong to the universality class of the
stable fixed point of the continuum AH field theory, which predicts a
continuous transition only for a large number of components, and in
particular in the large-$N$ limit.  Our numerical results provide a
bound on the number $N_c$ defined in Sec.~\ref{epsexp}, which
separates the small-$N$ first-order transition regime from the
large-$N$ continuous transition regime predicted by the 3D AH field
theory.  We obtain the upper bound $N_c<10$.  If we further assume
that the absence of continuous transitions along the CH transition
line of the model considered in this paper corresponds to the absence
of stable fixed points of the continuum AH field theory, we may
conclude that $N_c$ belongs to the interval $4 < N_c < 10$.

\noindent
(ii) For $N=2$ the CM transitions separating the Coulomb and molecular
phase are continuous and belong to the O(3) vector universality class,
as predicted by the LGW theory (\ref{hlg}), for sufficiently small
values of the inverse gauge coupling $\kappa$.  As $\kappa$ increases
along the CM line, the transition should eventually become of first
order, as all transitions are expected to be of first order close to
the multicritical point.  For $N\ge 3$, the CM transitions are
expected to be of first order, as predicted by the LGW theory.
 
\noindent
(iii) The transitions along the MH line are expected to be continuous,
and to belong to the XY universality class, at least for sufficiently
large values of the parameter $J$. However, we have not presented
results along this transition line.

The identification of the large-$N$ continuous CH transitions with the
universality class of the stable fixed point of the AH model is
strongly supported by the excellent agreement of the numerical results
for the critical exponent $\nu$ and $\eta$ with the predictions
obtained using the $1/N$ expansion within the continuum AH model. For
instance, for $N=25$ we find $\nu = 0.802(8)$ and $\eta_q = 0.883(7)$,
to be compared with the large-$N$ estimates $\nu = 0.805$ and $\eta_q
= 0.870$.  As far as we know, this is the first quantitative evidence
of such a correspondence.

It is worth comparing these results with those reported in
Ref.~\cite{PV-20-mfcp} for a lattice CP$^{N-1}$ model without
monopoles.  Numerical results for the monopole-free CP$^{N-1}$ model
provided clear evidence of a continuous transition for
$N=25$~\cite{PV-20-largeN}, at variance with what happens in the
standard CP$^{N-1}$, where the transition is discontinuous for any
$N\ge 3$.  Ref.~\cite{PV-20-mfcp} conjectured that the transition for
$N=25$ in the monopole-free model might be associated with the
large-$N$ stable fixed point of the continuum AH field theory. The
results of the present paper rule out this conjecture.  The estimate
of the critical exponent~\cite{PV-20-mfcp} $\nu=0.595(15)$ for the
$N=25$ monopole-free model definitely disagrees with the result
$\nu=0.802(8)$ obtained for the $N=25$ noncompact lattice AH model,
which is instead in agreement with the large-$N$ expansion of the
continuum AH field theory. Thus, we conclude that transitions of the
large-$N$ monopole-free CP$^{N-1}$ model are not described by the
continuum AH field theory.  It is tempting to conjecture that the
reason of the difference is in the nature of the coexisting phases at
the transition. The AH field theory is appropriate to describe
transitions between a Coulomb and a Higgs phase, but it is not
appropriate to describe the transition in the monopole-free CP$^{N-1}$
model. Indeed, in the latter case no Higgs phase exists: a disordered
monopole-free high-temperature phase coexists with a molecular phase
in which electromagnetic modes are still ungapped.
\bigskip

\emph{Acknowledgement}.  Numerical simulations have been performed on
the CSN4 cluster of the Scientific Computing Center at INFN-PISA.

\appendix 

\section{$C^*$ boundary conditions}
\label{appcstar}

We wish now to discuss the role that boundary conditions play in
noncompact formulations. We consider a finite system of size $L$ in
all directions.  As already discussed in Sec.~\ref{models}, the
partition function defined in Eq.~(\ref{ncz}) is ill-defined: $Z =
\infty$ for any $L$, because of gauge invariance. A standard way out
consists in considering only gauge-invariant observables and in
introducing a gauge fixing that eliminates all zero modes.  Let us
indicate symbolically with $G[A_{x,\mu}] = 0$ a maximal gauge fixing:
if $\{A_{x,\mu} \}$ is a configuration that satisfies the gauge-fixing
condition, there is no gauge transformation such the gauge-transformed
configuration also satisfies the gauge-fixing condition. Considering a
gauge invariant operator $B$, one can hope to obtain a well-defined
average value by defining
\begin{equation}
\langle B \rangle = 
  {\sum_{Az} B\, \delta(G)\, e^{-\beta H} \over \sum_{Az} \delta(G)\, e^{-\beta
H}} \; .
\label{Bgaugefixed}
\end{equation}
Unfortunately, in the case of periodic boundary conditions, also this
expression in ill-defined.

To clarify this issue, let us first consider the gauge Hamiltonian
$H_g$, Eq.~\eqref{nchg}.  It is invariant under the local gauge
transformation
\begin{equation}
A^{[\alpha]}_{{\bm r}, \mu} = A_{{\bm r}, \mu} + \alpha({\bm r}+\hat{\mu}) - 
       \alpha({\bm r})\ ,
\end{equation}
where $\alpha({\bm r})$ is an arbitrary function satisfying periodic
boundary conditions. The Hamiltonian $H_g$, however, is also invariant
under the shift
\begin{equation}
A_{{\bm r},\mu} \to A_{{\bm r},\mu}+c_{\mu}\ ,
\label{shiftsym}
\end{equation} 
where $c_{\mu}$ is a direction-dependent constant.  To clarify the
role played by the shift (\ref{shiftsym}), it is convenient to
introduce the noncompact Polyakov loop along the direction $\mu$,
defined by
\begin{equation}
P_{{\bm r_{\perp}},\mu}=\sum_{i=1}^{L}A_{(i,{\bm r_{\perp}}),\mu}\ .
\end{equation}
A generic point ${\bm r}$ is denoted with $(i,{\bm r_{\perp}})$ where
$r_{\mu}=i$, and ${\bm r_{\perp}}$ stands for the components of ${\bm
  r}$ different from the $\mu$-th one. It is immediate to verify that
the noncompact Polyakov loop is gauge invariant, while
\begin{equation}
P_{{\bm r_{\perp}},\mu}\to P_{{\bm r_{\perp}},\mu}+Lc_{\mu}
\end{equation}
under the transformation \eqref{shiftsym}. This shows that the shift
transformation cannot be rewritten as a gauge transformation.  As a
consequence there are three zero modes that cannot be eliminated by
the introduction of a gauge fixing. Therefore, in the absence of the
spin variables, also Eq.~(\ref{Bgaugefixed}) is ill-defined.

The transformation \eqref{shiftsym} is also present in compact
formulations. It corresponds to $\lambda_{{\bm x},\mu} \to
\lambda_{{\bm x},\mu} e^{i c_\mu}$.  However, in this case the
integration domain is compact and, therefore, zero modes do not make
average values ill-defined. This is obviously also the case of gauge
transformations and, indeed, in the compact case no gauge fixing is
needed to define rigorously the model.

The shift symmetry is broken when the spin fields $z_{{\bm x},\mu}$
are added.  However, because the gauge coupling of the spins is
obtained through the fields $\lambda_{{\bm x},\mu}$, transformations
such that $e^{i c_\mu} = 1$, leave the full Hamiltonian invariant.
Therefore, the infinite discrete subgroup of transformations
\begin{equation}
A_{{\bm r},\mu} \to A_{{\bm r},\mu}+2\pi n_{\mu}\ ,\qquad n_{\mu}\in\mathbb{Z}\,,
\label{shiftsym2}
\end{equation} 
is an invariance of the model, making expressions like
Eq.(\ref{Bgaugefixed}) ill-defined. An identical problem is
encountered in lattice Quantum Chromodynamics when studying the
electromagnetic properties of hadrons using a noncompact formulation
for the photon field (see, e.g., Refs.~\cite{LQED1, LQED2}).

To solve the problems mentioned above, we now discuss the $C^*$ boundary
conditions proposed in Ref.~\cite{LPRT-16}.  For the system studied in this
work, the $C^*$ boundary conditions are defined by the relations
\begin{equation} 
\label{cstardef}
A_{{\bm r} + L \hat{\nu}, \mu} = - A_{{\bm r}, \mu}\ ,  \qquad
{\bm z}_{{\bm r} + L \hat{\nu}} = \bar{\bm z}_{\bm r}\ .
\end{equation}
For consistency with Eq.~\eqref{cstardef}, the function $\alpha({\bm r})$
entering local gauge transformations
\begin{equation}
\begin{aligned}\label{gtran}
A^{[\alpha]}_{{\bm r}, \mu} &= A_{{\bm r}, \mu} + \alpha({\bm r}+\hat{\mu}) - 
       \alpha({\bm r})  \\
{\bm z}^{[\alpha]}_{\bm r} &= \exp[-i \alpha({\bm r})] {\bm z}_{\bm r}
\end{aligned}
\end{equation}
has to satisfy antiperiodic boundary conditions
\begin{equation}
\alpha({\bm r} + L \hat{\nu}) = - \alpha ({\bm r})\, .
\end{equation}
Moreover, from the relation ${\bm z}_{{\bm r} + L \hat{\nu}} = \bar{\bm z}_{\bm
r}$ it follows that the global U(1) symmetry is explicitly broken down to its
${\mathbb Z}_2$ subgroup:
\begin{equation}
A^{[\alpha]}_{{\bm r}, \mu} = A_{{\bm r}, \mu}\,, \qquad 
{\bm z}^{[\alpha]}_{\bm r} = s {\bm z}_{\bm r}\, ,
\end{equation}
with $s=\pm 1$.  Note that $C^*$ boundary conditions do not break
translational invariance, but care should be taken when performing
Fourier transforms.  For instance, the field $Q^{ab}$ is periodic,
while the plaquette operator is antiperiodic.

We will now show how to rewrite the previous conditions using only the
fields that belong to the cubic lattice $[1,L]^3$, changing the form
of the Hamiltonian for the sites and links close to the boundary.
This is necessary for the MC implementation.  Let us first consider
the gauge transformations: the transformation law of the scalar fields
in \eqref{gtran} does not require any modification, just like the
transformation rule of the gauge field when ${\bm r}+\hat{\mu} \in
[1,L]^3$. The gauge transformation of the fields $A_{{\bm r},\mu}$ on
the boundary of the cube can instead be rewritten, using the
anti-periodicity of $\alpha({\bm r})$, in the form
\begin{equation}
\begin{aligned}
A^{[\alpha]}_1(L,a,b) &= A_1(L,a,b) - \alpha(1,a,b) - \alpha(L,a,b)\\ 
A^{[\alpha]}_2(a,L,b) &= A_2(a,L,b) - \alpha(a,1,b) - \alpha(a,L,b)\\
A^{[\alpha]}_3(a,b,L) &= A_3(a,b,L) - \alpha(a,b,1) - \alpha(a,b,L) 
\end{aligned}
\end{equation}
where $a, b\in [1,L]$. 

The interaction term $H_z$ in Eq.~(\ref{hz}) is written as a sum of
terms which, for sites in the bulk of the lattice, are proportional to
\begin{equation}
h_{\mu}({\bm r})=\bar{\bm z}({\bm r}) \lambda_\mu({\bm r}) {\bm
  z}({\bm r} + \hat{\mu}) + \hbox{c.c.}\,.
\end{equation}
For  sites on the boundary of the lattice instead, the interactions can 
be written as 
\begin{equation}
\begin{aligned}
h_1(L,a,b) &=\bar{\bm z}(L,a,b) \lambda_1(L,a,b) \bar{\bm z}(1,a,b) + 
    \hbox{c.c.} \\
h_2(a,L,b) &= \bar{\bm z}(a,L,b) \lambda_2(a,L,b) \bar{\bm z}(a,1,b) + 
    \hbox{c.c.} \\
h_3(a,b,L) &= \bar{\bm z}(a,b,L) \lambda_3(a,b,L) \bar{\bm z}(a,b,1) + 
    \hbox{c.c.} 
\end{aligned}
\end{equation}
These are the terms that explicitly break the global U(1) invariance.
They are still gauge invariant, due to the different gauge
transformations that are applied on the field $A_{{\bm r},\mu}$ on the
boundary.

The noncompact gauge interaction term $H_g$ in \eqref{nchg} is written as a sum
of terms involving the noncompact plaquette operator, which for sites in the
bulk of the lattice can be written as
\begin{equation}
\Pi_{\mu\nu}({\bm r}) = \left[
   A_\mu({\bm r}) + A_\nu({\bm r} + \hat{\mu}) 
   -A_\mu({\bm r} + \hat{\nu}) - A_\nu({\bm r}) \right]^2\,.
\end{equation}
For plaquettes on the boundaries this expression has to be changed and
we provide here explicit expressions for the case case $(\mu,\nu) =
(1,2)$ (the other two cases are completely analogous). For $a\in
[1,L)$ and $b\in [1, L]$ we have to use
\begin{equation}
\begin{aligned}
\Pi_{12}(L,a,b) &= \left[
    A_1(L,a,b) - A_2(1,a,b) \right. \\
    &\quad -\left.
   A_1(L,a+1,b) - A_2(L,a,b) \right]^2 \\
\Pi_{12}(a,L,b) &= \left[
    A_1(a,L,b) + A_2(a+1,L,b) \right. \\
   &\quad + \left. A_1(a,1,b) - A_2(a,L,b) \right]^2 \\
\Pi_{12}(L,L,b) &= \left[
    A_1(L,L,b) - A_2(1,L,b) \right. \\
   &\quad + \left. A_1(L,1,b) - A_2(L,L,b) \right]^2 \ ,
\end{aligned}
\end{equation}
which are easily shown to be gauge invariant.

Let us now show that $C^*$ boundary conditions eliminate the shift
symmetry that makes periodic boundary conditions ill-defined.  Indeed,
in the $C^*$ case, Polyakov loops are not gauge invariant. Using for
definiteness the Polyakov loop in the $\hat{z}$ direction, i.e.,
\begin{equation}
P_3(x, y) = \sum_{z=1}^L A_3(x,y,z) , 
\end{equation}
we have 
\begin{equation}
P_3^{[\alpha]}(x,y) = P_3(x, y) - 2 \alpha(x,y,1)\ .
\end{equation}
A simple consequence of this fact is that, by means of local gauge
transformations, we can enforce $A_3({\bm r}) = 0$ for all points,
obtaining the maximal temporal gauge (this is obviously not possible
when using periodic boundary conditions, since Polyakov loops are
gauge invariant in that case).  The algorithm to implement the maximal
temporal gauge is the following. We first perform a gauge
transformation with $\alpha(x,y,2) = -A_3(x,y,1)$ and $\alpha(x,y,z) =
0$ for $z\not = 2$ (this fixes $A_3(x,y,1)=0$), then we use a gauge
transformation with $\alpha(x,y,3) = -A_3(x,y,2)$ and $\alpha(x,y,z) =
0$ for $z\not = 3$ and so on, until we reach the plane $z=L$.  At this
point only $A_3(x,y,L)$ is not vanishing and we perform a
transformation with $\alpha(x,y,z) = A_3(x,y,L)/2$ (the same for all
$z$ values).

To conclude the proof that $C^*$ boundary conditions make the
gauge-fixed theory well-defined, let us show that, once the maximal
temporal gauge is introduced, there is a unique minimum of the gauge
Hamiltonian $H_g$, confirming the absence of dangerous zero modes.
Starting from
\begin{equation}
\Pi_{13}(x,y,1) = [A_1(x,y,1) - A_1(x,y,2)]^2 
\end{equation}
we obtain by minimization $A_1(x,y,2) = A_1(x,y,1)$ for all $x,y$.  If
we now consider $\Pi_{13}(x,y,2)$, we obtain in the same way
$A_1(x,y,3) = A_1(x,y,2)$, and repeating the same procedure for
$\Pi_{13}(x,y,z)$ with $1\le z \le L-1$, we get $A_1(x,y,z) =
A_1(x,y,1)$ for all $x,y,z$.  The minimization of the boundary
plaquette
\begin{equation}
\begin{aligned}
\Pi_{13}(x,y,L) &= [A_1(x,y,1) + A_1(x,y,L)]^2 \\
                &= 4A_1(x,y,1)^2
\end{aligned}
\end{equation}
finally implies $A_1(x,y,1) = 0$, hence $A_1(x,y,z) = 0$ for all
$x,y,z$.  Using $\Pi_{23}$ instead of $\Pi_{13}$ the same argument
shows that $A_2(x,y,z) = 0$. We have therefore proved that, at
variance with the case of periodic boundary conditions, when using
$C^*$ boundary conditions, there is a single configuration (up to
gauge transformations) that minimize the gauge Hamiltonian: no
gauge-invariant zero modes are present.

To conclude the appendix, let us go back to the question of the gauge
fixing.  We have proved that the statistical averages are well defined
provided that $C^*$ boundary conditions and a maximal gauge fixing are
used.  However, in the simulation we have not introduced any gauge
fixing. We wish now to explain why the gauge fixing is irrelevant in
MC calculations of gauge-invariant observables. Let us collectively
call $\phi_t$ the fields we have generated at MC time $t$. There is
obviously a gauge transformation that maps $\phi_t$ onto new fields
$\phi^{[\alpha]}_t$ that satisfy the gauge fixing condition: the
correspondence between $\phi_t$ and $\phi^{[\alpha]}_t$ is
unique. Therefore, the dynamics $\phi_1\to \phi_2 \to \phi_3 \ldots$
can be mapped onto the dynamics $\phi^{[\alpha]}_1 \to
\phi^{[\alpha]}_2 \to \phi^{[\alpha]}_3 \ldots$. Thus, even if we do
not implement the gauge-fixing condition, gauge-invariant quantities
take the same values as if they were obtained in a simulation in the
gauge-fixed model.  This is, of course, not true for
non-gauge-invariant quantities: for instance, the fields $A_{{\bm
    x},\mu}$ perform a random walk and their absolute values increase
with time: their averages are not defined.

\end{document}